\documentclass[aps,prl,twocolumn,preprintnumbers,amsmath,amssymb,superscriptaddress,10pt]{revtex4-2}

\usepackage[utf8]{inputenc}
\usepackage{graphicx}
\usepackage{bm}

\usepackage{color}
\definecolor{darkblue}{rgb}{0,0,0.6}
\definecolor{red}{rgb}{1,0,0}
\definecolor{blue}{rgb}{0,0,1}

\usepackage{braket}

\newcommand{\ou}{%
  \mathrel{%
    \vcenter{\offinterlineskip
      \ialign{##\cr$<$\cr\noalign{\kern-1.5pt}$>$\cr}%
    }%
  }%
}

\def\c{\ensuremath \hat{c}}

\usepackage{ifpdf}
\ifpdf
\usepackage{epstopdf}
\usepackage[pdftex,unicode,pdfstartview={FitH},pdfborder={0 0 0}]{hyperref}
\usepackage{hypcap}
\else
\usepackage[hypertex]{hyperref}
\fi
\hypersetup{
    bookmarksnumbered = true,
    colorlinks = true, linkcolor = darkblue,
    citecolor = darkblue, filecolor = darkblue,
    menucolor = darkblue, urlcolor = darkblue
}

\usepackage{dcolumn}
\newcolumntype{R}{>{$\displaystyle}r<{$}}
\newcolumntype{C}{>{$\displaystyle}c<{$}}
\newcolumntype{L}{>{$\displaystyle}l<{$}}

\begin{document}
    
\title{Implementation of the bilayer Hubbard model in a moir\'e heterostructure}

\author{Borislav Polovnikov*}
\affiliation{Fakult\"at f\"ur Physik, Munich Quantum Center, Ludwig-Maximilians-Universit\"at M\"unchen, Geschwister-Scholl-Platz 1, 80539 M\"unchen, Germany}
\affiliation{Center for NanoScience (CeNS), Ludwig-Maximilians-Universit\"at M\"unchen, Geschwister-Scholl-Platz 1, 80539 M\"unchen, Germany}
\author{Johannes Scherzer*}
\affiliation{Fakult\"at f\"ur Physik, Munich Quantum Center, Ludwig-Maximilians-Universit\"at M\"unchen, Geschwister-Scholl-Platz 1, 80539 M\"unchen, Germany}
\affiliation{Center for NanoScience (CeNS), Ludwig-Maximilians-Universit\"at M\"unchen, Geschwister-Scholl-Platz 1, 80539 M\"unchen, Germany}
\author{Subhradeep Misra*}
\affiliation{Fakult\"at f\"ur Physik, Munich Quantum Center, Ludwig-Maximilians-Universit\"at M\"unchen, Geschwister-Scholl-Platz 1, 80539 M\"unchen, Germany}
\affiliation{Center for NanoScience (CeNS), Ludwig-Maximilians-Universit\"at M\"unchen, Geschwister-Scholl-Platz 1, 80539 M\"unchen, Germany}
\author{Henning Schl\"omer}
\affiliation{Fakult\"at f\"ur Physik, Munich Quantum Center, Ludwig-Maximilians-Universit\"at M\"unchen, Geschwister-Scholl-Platz 1, 80539 M\"unchen, Germany}
\affiliation{Arnold Sommerfeld Center,  Ludwig-Maximilians-Universit\"at M\"unchen, Theresienstrasse 37, 80333 M\"unchen, Germany}
\author{Julian Trapp}
\affiliation{Fakult\"at f\"ur Physik, Munich Quantum Center, Ludwig-Maximilians-Universit\"at M\"unchen, Geschwister-Scholl-Platz 1, 80539 M\"unchen, Germany}
\affiliation{Center for NanoScience (CeNS), Ludwig-Maximilians-Universit\"at M\"unchen, Geschwister-Scholl-Platz 1, 80539 M\"unchen, Germany}
\author{Xin Huang}
\affiliation{Beijing National Laboratory for Condensed Matter Physics, Institute of Physics, Chinese Academy of Sciences, Beijing 100190, People’s Republic of China}
\affiliation{School of Physical Sciences, CAS Key Laboratory of Vacuum Physics, University of Chinese Academy of Sciences, Beijing 100190, People’s Republic of China}
\author{Christian Mohl}
\affiliation{Fakult\"at f\"ur Physik, Munich Quantum Center, Ludwig-Maximilians-Universit\"at M\"unchen, Geschwister-Scholl-Platz 1, 80539 M\"unchen, Germany}
\affiliation{Center for NanoScience (CeNS), Ludwig-Maximilians-Universit\"at M\"unchen, Geschwister-Scholl-Platz 1, 80539 M\"unchen, Germany}
\author{Zhijie Li}
\affiliation{Fakult\"at f\"ur Physik, Munich Quantum Center, Ludwig-Maximilians-Universit\"at M\"unchen, Geschwister-Scholl-Platz 1, 80539 M\"unchen, Germany}
\affiliation{Center for NanoScience (CeNS), Ludwig-Maximilians-Universit\"at M\"unchen, Geschwister-Scholl-Platz 1, 80539 M\"unchen, Germany}
\author{Jonas G{\"o}ser}
\affiliation{Fakult\"at f\"ur Physik, Munich Quantum Center, Ludwig-Maximilians-Universit\"at M\"unchen, Geschwister-Scholl-Platz 1, 80539 M\"unchen, Germany}
\affiliation{Center for NanoScience (CeNS), Ludwig-Maximilians-Universit\"at M\"unchen, Geschwister-Scholl-Platz 1, 80539 M\"unchen, Germany}
\author{Jonathan F\"orste}
\affiliation{Fakult\"at f\"ur Physik, Munich Quantum Center, Ludwig-Maximilians-Universit\"at M\"unchen, Geschwister-Scholl-Platz 1, 80539 M\"unchen, Germany}
\affiliation{Center for NanoScience (CeNS), Ludwig-Maximilians-Universit\"at M\"unchen, Geschwister-Scholl-Platz 1, 80539 M\"unchen, Germany}
\author{Ismail Bilgin}
\affiliation{Fakult\"at f\"ur Physik, Munich Quantum Center, Ludwig-Maximilians-Universit\"at M\"unchen, Geschwister-Scholl-Platz 1, 80539 M\"unchen, Germany}
\affiliation{Center for NanoScience (CeNS), Ludwig-Maximilians-Universit\"at M\"unchen, Geschwister-Scholl-Platz 1, 80539 M\"unchen, Germany}
\author{Kenji Watanabe}
\affiliation{Research Center for Electronic and Optical Materials, National Institute for Materials Science, 1-1 Namiki, Tsukuba 305-0044, Japan}
\author{Takashi Taniguchi}
\affiliation{International Center for Materials Nanoarchitectonics, National Institute for Materials Science, 1-1 Namiki, Tsukuba 305-0044, Japan}
\author{Annabelle Bohrdt}
\affiliation{Munich Center for Quantum Science and Technology (MCQST), Schellingstra\ss{}e 4, 80799 M\"unchen, Germany}
\affiliation{Institut f\"ur theoretische Physik, Universit\"at Regensburg, 93035 Regensburg, Germany}
\author{Fabian Grusdt}
\affiliation{Fakult\"at f\"ur Physik, Munich Quantum Center, Ludwig-Maximilians-Universit\"at M\"unchen, Geschwister-Scholl-Platz 1, 80539 M\"unchen, Germany}
\affiliation{Arnold Sommerfeld Center,  Ludwig-Maximilians-Universit\"at M\"unchen, Theresienstrasse 37, 80333 M\"unchen, Germany}
\affiliation{Munich Center for Quantum Science and Technology (MCQST), Schellingstra\ss{}e 4, 80799 M\"unchen, Germany}
\author{Anvar~S.~Baimuratov}
\affiliation{Fakult\"at f\"ur Physik, Munich Quantum Center, Ludwig-Maximilians-Universit\"at M\"unchen, Geschwister-Scholl-Platz 1, 80539 M\"unchen, Germany}
\affiliation{Center for NanoScience (CeNS), Ludwig-Maximilians-Universit\"at M\"unchen, Geschwister-Scholl-Platz 1, 80539 M\"unchen, Germany}
\author{Alexander H{\"o}gele}
\affiliation{Fakult\"at f\"ur Physik, Munich Quantum Center, Ludwig-Maximilians-Universit\"at M\"unchen, Geschwister-Scholl-Platz 1, 80539 M\"unchen, Germany}
\affiliation{Center for NanoScience (CeNS), Ludwig-Maximilians-Universit\"at M\"unchen, Geschwister-Scholl-Platz 1, 80539 M\"unchen, Germany}
\affiliation{Munich Center for Quantum Science and Technology (MCQST), Schellingstra\ss{}e 4, 80799 M\"unchen, Germany}

\maketitle

\textbf{Moir\'e materials provide a unique platform for studies of correlated many-body physics of the Fermi-Hubbard model on triangular spin-charge lattices. Bilayer Hubbard models are of particular significance with regard to the physics of Mott insulating states and their relation to unconventional superconductivity, yet their experimental implementation in moir\'e systems has so far remained elusive. Here, we demonstrate the realization of a staggered bilayer triangular lattice of electrons in an antiparallel MoSe$_{2}$/WS$_{2}$ heterostructure. The bilayer lattice emerges due to strong electron confinement in the moir\'e potential minima and the near-resonant alignment of conduction band edges in MoSe$_{2}$ and WS$_{2}$. As a result, charge filling proceeds layer-by-layer, with the first and second electron per moir\'e cell consecutively occupying first the MoSe$_{2}$ and then the WS$_{2}$ layer. We describe the observed charging sequence by an electrostatic model and provide experimental evidence of spin correlations on the vertically offset and laterally staggered bilayer lattice, yielding absolute exciton Land\'e factors as high as $600$ at lowest temperatures. The bilayer character of the implemented spin-charge lattice allows for electrostatic tunability of Ruderman-Kittel-Kasuya-Yosida magnetism, and establishes antiparallel MoSe$_{2}$/WS$_{2}$ heterostructures as a viable platform for studies of bilayer Hubbard model physics with exotic magnetic phases on frustrated lattices.}

The two-dimensional Hubbard model is paradigmatic in solid state physics, with theoretically described phases ranging from Mott insulating states, through quantum spin-liquids~\cite{Balents2010Mar} to superconductivity in the strongly coupled $t-J$ limit~\cite{Lee2006jan}. Motivated by the enigmatic relationship between the number of CuO$_2$ layers and the manifestation of high-temperature superconductivity in the cuprate family~\cite{Iwano2022Aug, Craco2022Aug}, the Hubbard model has received extensive theoretical consideration on a bilayer lattice~\cite{Hetzel1994Aug,dosSantos1995Jun,Maier2011Nov,Bohrdt2021Dec,Bohrdt2022Jun} or in the presence of an incipient band~\cite{Kato2020May}. In a triangular lattice geometry~\cite{Venderley2019Aug, Zhu2022May,Mou2022Sep}, however, strong ground state degeneracy caused by geometric frustration, together with the notorious sign problem, render theoretical calculations challenging. 

While cold-atom quantum simulation of the Hubbard model based on quantum gas microscopes can provide insight into different phases on a square-lattice geometry~\cite{Chiu2019Jul,Hirthe2023Jan,Mazurenko2017May,Gall2021Jan,BlochGreiner2022,Bohrdt2021Dec}, heterostructures of transition metal dichalcogenides (TMDs) have emerged as an increasingly important platform for quantum simulation on triangular lattices~\cite{Tang2020,Wang2020a,Gerardot_exciton-polarons_2022,Tang2023Mar,Morera2023Jun,Ciorciaro2023May}. The geometric moir\'e interference effect in stacked van der Waals heterostructures plays a pivotal role in this development, since critical parameters of the long-range moir\'e potential can be engineered via the constituent layer materials and their rotational alignment. Moreover, following the estimate of the spin-exchange coupling in TMD heterobilayers (HBLs) of $J=0.05$~meV~\cite{Tang2020}, the temperature range accessible with modern dilution refrigerators yields energy scales as low as $T/J \lesssim 1/10$, paving the way for systematic studies of frustrated magnetic order in two dimensions~\cite{Tang2020,Gerardot_exciton-polarons_2022,Tang2023Mar,Morera2023Jun,Ciorciaro2023May}.

In this work, we establish antiparallel MoSe$_2$/WS$_2$ heterostructures as a two-dimensional bilayer Hubbard model system with triangular geometry. Using low-temperature optical spectroscopy, we demonstrate that strong electron confinement in moir\'e potential minima opens an energy gap of about $60$~meV. This gap in turn gives rise to discretization of the density of states (DOS) and leads to a peculiar charging behavior: in a first step, the primary lattice in the MoSe$_2$ layer is charged with one electron per moir\'e cell, and in a second step, the electrons fill a vertically offset and laterally staggered secondary lattice in the WS$_2$ layer. This compound charge lattice implements a staggered bilayer Hubbard model in the strong interaction limit with stabilized magnetic correlations in a wide range of electrostatic gate voltages. We perform measurements of doping-dependent spin susceptibility that indicate antiferromagnetic exchange interactions and, in addition, suggest the presence of Ruderman-Kittel-Kasuya-Yosida (RKKY) magnetism above doping levels of one electron per moir\'e cell. 

\begin{figure}[t!]  
\includegraphics[scale=0.92]{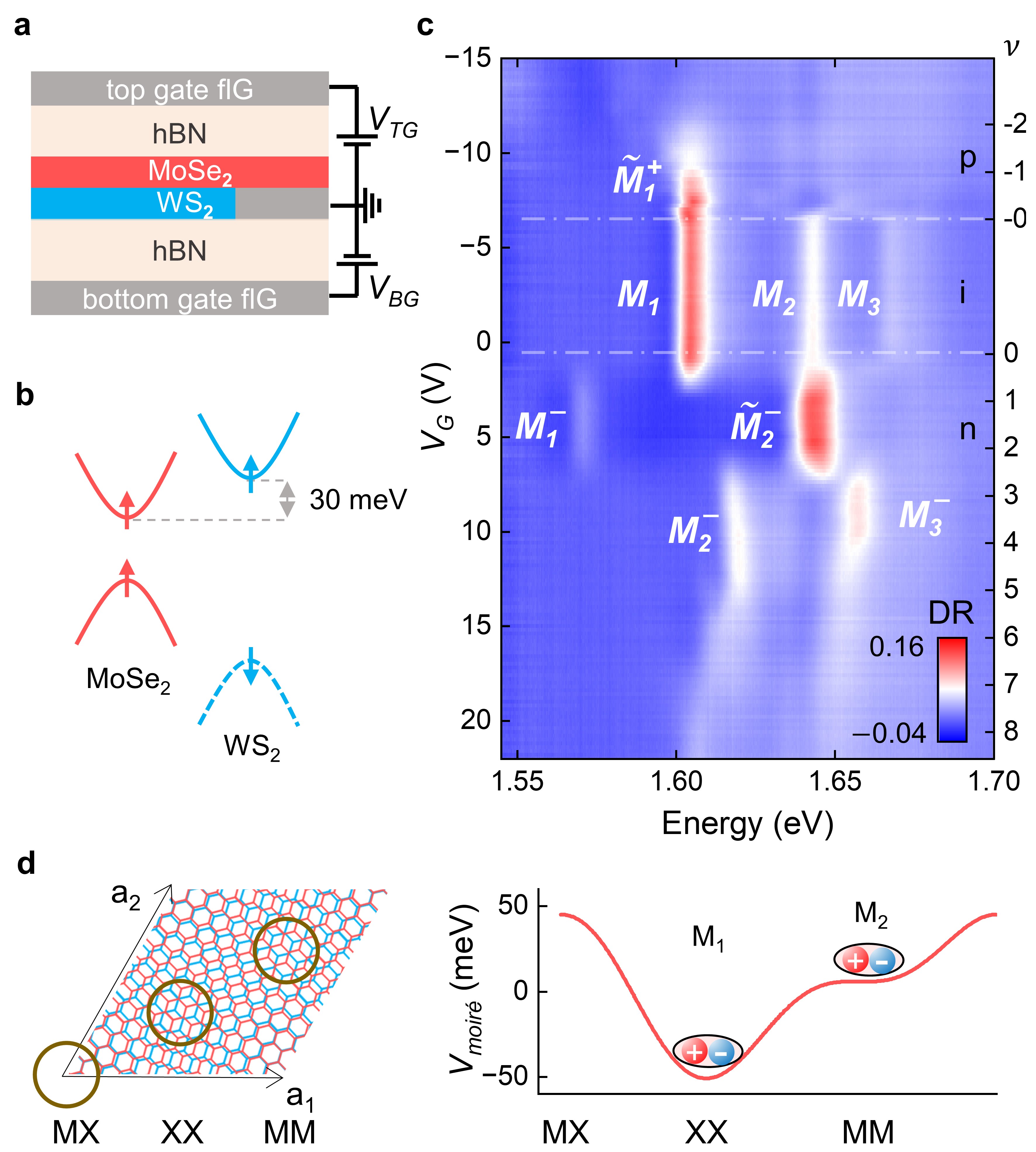}
\caption{\textbf{Charging characteristics of antiparallel MoSe$_{2}$/WS$_{2}$.} \textbf{a}, Device layout with top and bottom few-layer graphite gates, hBN dielectric layers, and the MoSe$_{2}$/WS$_{2}$ HBL sandwiched in between. \textbf{b}, Type I alignment of conduction and valence bands in antiparallel stacking, with arrows indicating co-polarized spins in $K$ and $K'$ valleys of MoSe$_{2}$ and WS$_{2}$, respectively. \textbf{c}, Evolution of the DR spectra as a function of symmetrically applied gate voltages for both hole (p) and electron (n) doping. Neutral moir\'e excitons $M_1$, $M_2$, and $M_3$ show different responses to doping, consistent with different spatial localization in the moir\'e cell. \textbf{d}, Schematics of the moir\'e lattice geometry with high-symmetry points inside the unit cell (left) and the moir\'e exciton potential (right).}
\label{fig1}
\end{figure}

The effect is identified in field-effect devices consisting of the MoSe$_2$/WS$_2$ HBL encapsulated in symmetric hexagonal boron nitride (hBN) dielectric spacers and sandwiched between top and bottom few-layer graphene gates. One sample (S1) was assembled from monolayers grown by chemical vapor deposition, and another (S2) was fabricated from exfoliated monolayers (see Supplementary Information). Similar to WS$_2$/WSe$_2$ heterostacks, MoSe$_2$/WS$_2$ HBLs feature a relatively large lattice mismatch of 4\%, which impedes mesoscopic lattice reconstruction~\cite{Zhao2023} and stabilizes the canonical triangular moir\'e geometry shown in the left panel of Fig.~\ref{fig1}d. In this limit, the moir\'e pattern varies spatially through the points of high-symmetry registries MM, MX and XX~\cite{mak_semiconductor_2022} (with M and X representing metal and chalcogen atoms, respectively). The periodically modulated moir\'e potentials for electrons, holes and excitons~\cite{MacdonaldTopo2017,MacdonaldIX2018,MacdonaldHubbard2018,Tong2020Dec} give rise to distinct, energetically favored spatial positions, as illustrated in the right panel of Fig.~\ref{fig1}d. In particular, umklapp-scattering off the long-range superlattice gives rise to pronounced exciton mixing and the emergence of robust moir\'e excitons ~\cite{AlexeevTartakovsky2019,Zhang_Deng_Twist2020,Tang2021,Tang2022,Polovnikov2023Apr,Ciorciaro2023May}.

Figure \ref{fig1}c shows the evolution of the differential reflectance (DR) of sample S1 as a function of symmetric gate voltage, $V_G = V_{TG} = V_{BG}$. In the neutral regime from $0.6$ to $-7.0$~V, three bright moir\'e excitons are observed close to the energy of the MoSe$_2$ A-exciton. At the boundaries of the neutral region, optical signatures indicate transitions to both the electron ($0.6$~V) and the hole ($-7.0$~V) doped regimes. On the p-doped side, the lowest energy exciton $M_1$ at $1.60$~eV exhibits a series of step-like red and blue-shifts before losing its oscillator strength and giving rise to a faint positive trion $M_1^+$ with $25$~meV red-shift, whereas the higher-energy peaks $M_2$ and $M_3$ at $1.60$ and $1.67$~eV, respectively, disappear as soon as charge doping into the valence band sets on. On the n-doped side this behavior is reversed: in a first charging step, $M_1$ converts abruptly into a negative trion $M_1^-$ with a binding energy of $33$~meV, whereas $M_2$ evolves gradually into a slightly red-shifted peak $\tilde{M}_2^-$ before jumping abruptly to $M_2^-$ in a second charging step. In particular, this second transition at $6.5$~V coincides with the emergence of a resonance between $M_2$ and $M_3$ which we identify as the charged exciton $M_3^-$, and with a similarly abrupt quench of the ground state trion $M_1^-$. Finally, in a third charging step around $13.0$~V, both $M_2^-$ and $M_3^-$ red-shift and lose their oscillator strength.

Consistent with previous studies~\cite{FengWang2019,naik_intralayer_2022,Polovnikov2023Apr}, the contrasting responses of $M_1$ and $M_2$ to positive and negative charge doping can be attributed to distinct spatial positions of the two excitons within the moir\'e unit cell, as illustrated in Fig.~\ref{fig1}d. The n-doped side is particularly instructive: just as in the case of parallel alignment~\cite{Ciorciaro2023May}, the charged trion $M_1^-$ indicates that doping-induced electrons are co-localized with $M_1$ at the moir\'e potential minima of XX sites~\cite{Polovnikov2023Apr, atac_dft}. The second exciton $M_2$, on the contrary, was predicted to be located at the MM site, which in the limiting case of perfect rotational alignment corresponds to a lateral displacement of $\sim 4$~nm. Therefore, prior to the second doping transition, the exciton $\tilde{M}_2^-$ acts as a remote sensor~\cite{Xu2020}, with binding energy and oscillator strength acting as probes of the surrounding electron lattice.

\begin{figure*}[t!]  
\includegraphics[scale=0.97]{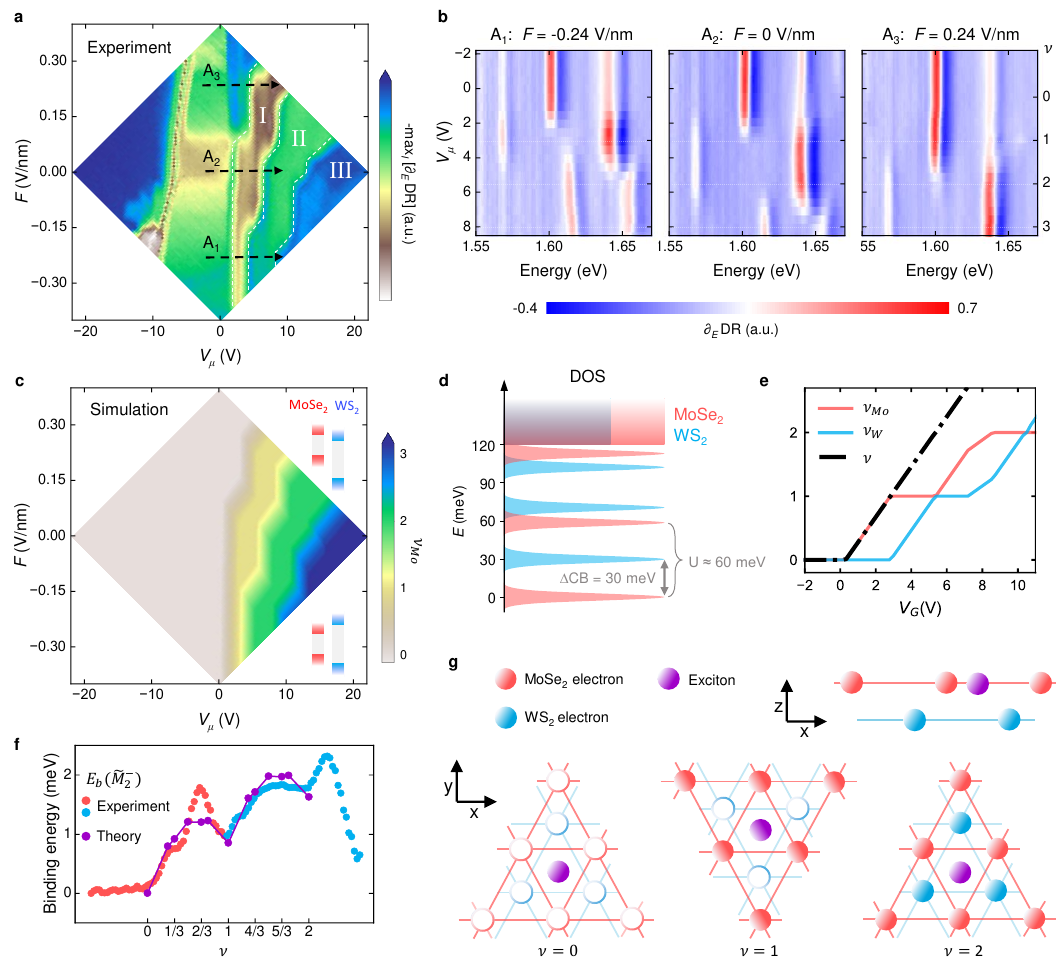}
\caption{\textbf{Charging behavior upon electron doping.} \textbf{a}, Hyperspectral DR map, where the color of each pixel represents the negative maximum of the derivative $\partial_E \text{DR}(E)$ in the interval between $1.603$ and $1.800$~eV to highlight the different charging states. The left side represents the p-doped regime, the central vertical stripe the intrinsic, and the right side the n-doped regime. The boundaries of the three distinct regions on the n-doped side (I, II and III) signify subsequent charging steps in the MoSe$_2$ layer. \textbf{b}, Line-cuts of the data in \textbf{a} for three representative electric fields. The onset of electron doping shows a crossover from type~I at negative and small positive fields (lines $A_1$ and $A_2$) to type~II band alignment for elevated positive fields (line $A_3$). Charged excitons $M_1^-$ and $\tilde{M}_2^-$ are present throughout the region I and reflect the charge distribution among the MoSe$_2$ and WS$_2$ layers. \textbf{c}, Simulation of the electron density in the MoSe$_2$ layer as a function of $V_\mu$ and $F$, reproducing the most pronounced features of electron doping. \textbf{d}, Schematics of the DOS in both layers obtained from the simulation. Doping of the layers proceeds in steps of $n_0$, and the step-like extent of the region I implies Coulomb repulsion of $60$~meV between the first and the second electron charging event in the MoSe$_2$ layer. Together with the CB offset of $30$~meV this leads to peculiar charging behavior shown in \textbf{e}, with charging of the MoSe$_2$ layer up to one electron per moir\'e cell and subsequent charge stability upon consecutive filling of the WS$_2$ layer up to the same filling factor. \textbf{f}, Experimental and theoretical binding energy of $\tilde{M}_2^-$, evaluated as the red-shift from the energy of its neutral counterpart $M_2$, as a function of the electron filling factor along the line $A_2$. \textbf{g}, Schematics of the moir\'e exciton position and sublattice charge ordering at integer filling factors.}
\label{fig2}
\end{figure*}

In the following analysis, we employ $\tilde{M}_2^-$ to obtain insight into doping characteristics at varying electric fields and establish evidence of stabilized spin-charge order in the MoSe$_2$ layer. To this end, we study the DR signal as a function of both top and bottom gate voltages $V_{TG}$ and $V_{BG}$. The charge density in the HBL is determined by the doping potential $V_\mu=(V_{TG}+V_{BG})/2$, whereas the electric field is given by $F=(V_{BG}-V_{TG})/l$, with $l=110$~nm being the total thickness of hBN layers. For aggregated visualization of the hyper-spectral data, we show for each point $(V_\mu , F)^T$ the negative maximum of the derivative $d(\text{DR})/dE$ between $1.603$ and $1.800$~eV in Fig.~\ref{fig2}a. Extended regions of constant color indicate constant optical response, whereas straight lines and kinks between these regions represent transitions mediated by charging. In particular, the three consecutive electron-doping regions denoted by I, II and III represent the three charging steps of MoSe$_2$ discussed above (see also Extended Data Fig.~\ref{SMfigSpectra}).

The extent of these three charge-stability regions varies strongly with the applied electric field. In Fig.~\ref{fig2}b, we show the evolution of $M_1$ and $M_2$ with $V_\mu$ for three distinct electric fields indicated by the three lines in Fig.~\ref{fig2}a. For $F=0.24$~V/nm (right panel), the region~I is shifted to higher voltages, with neutral excitons vanishing at $\sim 4$~V which is half-way through the first charging step at zero-field (central panel). The onset of this shift happens at a field of $F_0=0.1$~V/nm, which is consistent with a conduction band (CB) offset of $30$~meV between MoSe$_2$ and WS$_2$~\cite{Polovnikov2023Apr}. Thus, the behavior at fields higher than $F_0$ is readily explained: in this regime, the CB edge of WS$_2$ has been tuned below that of MoSe$_2$ through a crossover from type~I to type~II band alignment, forcing electrons into the WS$_2$ layer first. The red-shift of $M_1$, observed at $\sim 1.0$~V in the right panel of Fig.~\ref{fig2}b, confirms the presence of electrons, and hence a change of the dielectric environment in the WS$_2$ layer. The excitons $M_1$ and $M_2$ in the MoSe$_2$ layer~\cite{Polovnikov2023Apr,Tang2021,Tang2022} are unable to form intralayer charge-bound states until electrons start filling the MoSe$_2$ sublattice at higher $V_\mu$, and the emergence of $M_1^-$ and  $\tilde{M}_2^-$ is shifted to higher voltages accordingly.

For the negative field $F= -0.24$~V/nm (left panel), on the contrary, the onset of electron doping coincides with the zero-field case, but the width of the charging region~I with stability of both $M_1^-$ and $\tilde{M}_2^-$ is reduced by half. This behavior is more intriguing, since it indicates that for electric fields pointing from the MoSe$_2$ to the WS$_2$ layer, the second charging transition occurs earlier than at zero electric field. From a different perspective, it implies that the number of electrons added to the MoSe$_2$ layer along the zero-field line is reduced, resulting in an extended charge stability range of region~I. Notably, this charging behavior is absent in parallel MoSe$_2$/WS$_2$ HBLs, where the moir\'e potential is deeper and can thus accommodate more electrons within one layer (see Extended Data Fig.~\ref{SMfigRType}).

To explain this intricate charging behavior, we perform electrostatic simulations with a discretized quantum capacitance or DOS inside the two layers~\cite{Tan2023May,Popert2022}. Since both the geometric capacitance and the moir\'e density $n_0 \approx 2.0 \times 10^{12}\, \text{cm}^{-2}$ are fixed, the only free degree of freedom pertinent to the charge carrier density is the quantum capacitance, which quantifies the cost of electrochemical energy per charge carrier induced in the HBL, $n(E) = \int_{0}^E \text{DOS}(E') dE'$. We assume that carrier doping happens in steps of $n_0$ for both layers, with each step corresponding to a peak in the DOS as shown in Fig.~\ref{fig2}d. The gaps between these peaks represent energetic cost associated with on-site Coulomb repulsion $U$ due to strong confinement in moir\'e potential pockets. 

To model the charging behavior of the HBL, we adjust the energy of consecutive peaks with finite DOS to recover the same extents for the regions~I, II and III as in Fig.~\ref{fig2}a (see Methods and Extended Data Fig.~\ref{SMfigModel} for details). Figure~\ref{fig2}c shows the simulation result on the electron-doped side with very good agreement with experimental data. Remarkably, the simulation implies that region~I represents a regime with just one electron per moir\'e cell in MoSe$_2$, with excess electrons populating the WS$_2$ layer instead. Region~II, with the presence of the peaks $M_2^-$ and $M_3^-$, corresponds to two electrons inside MoSe$_2$, and the region~III is characterized by reduced oscillator strength of the slightly red-shifted $M_2^-$ and $M_3^-$ with three and more electrons per moir\'e cell inside the MoSe$_2$ layer.

\begin{figure*}[ht!]  
\includegraphics[scale=0.96]{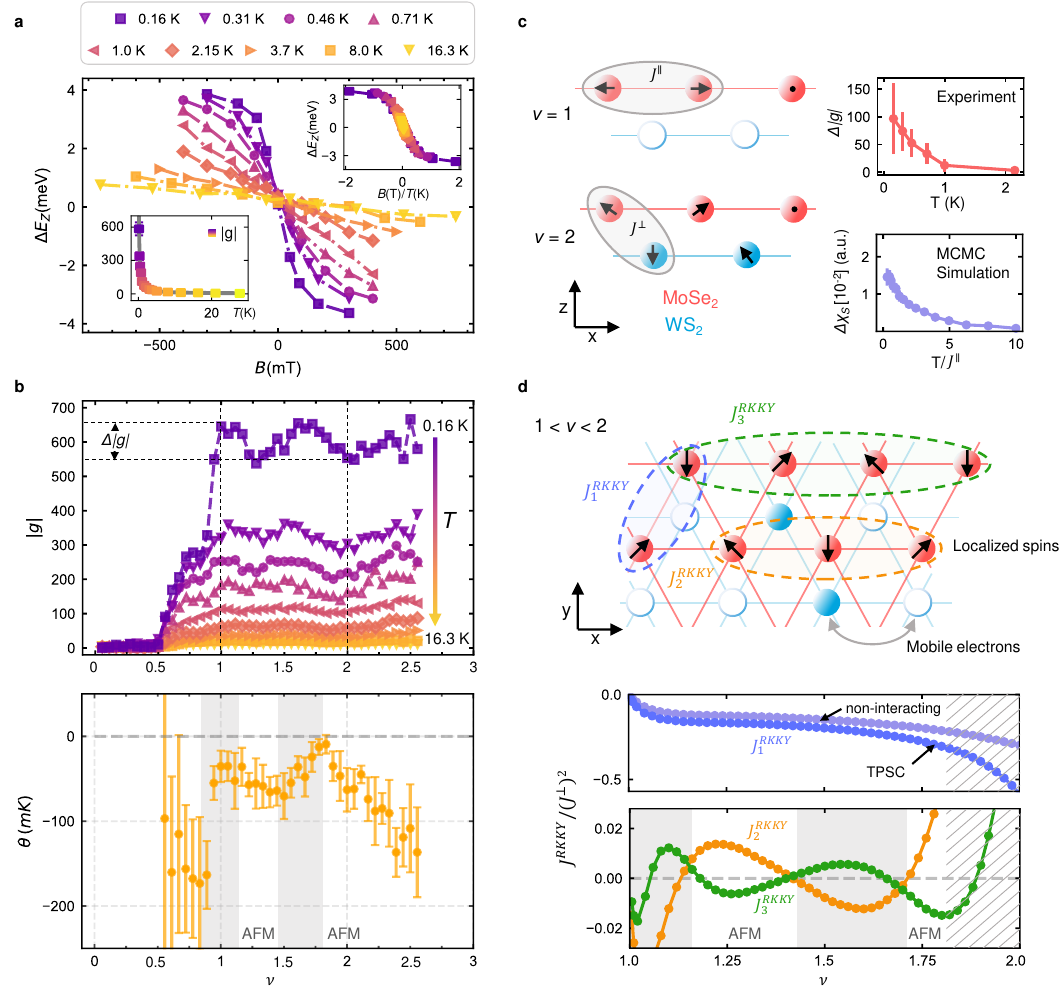}
\caption{\textbf{Correlated magnetism upon electron doping.} \textbf{a}, Zeeman splittings of $\tilde{M}_2^-$ in $\sigma^+$ and $\sigma^-$ polarization, with colors indicating experimental temperatures. Right inset: same data as a function of the rescaled field $B/T$; left inset: Curie-Weiss fit (solid line) to the temperature-dependence of $g$-factors. \textbf{b}, Upper panel: Magnitude of the $g$-factor as function the filling factor at different temperatures; the difference in $g$-factors at integer fillings $\nu=1$ and $2$, $\Delta |g|$, is indicated by the black dashed lines. Lower panel: Curie-Weiss temperature $\theta$ extracted from fits to the data in \textbf{a}. An increase (decrease) of $\theta$, highlighted by grey (white) areas, corresponds to a weakening (strengthening) of the $120^\circ$ antiferromagnetic (AFM) correlations. \textbf{c}, Left panel: Schematics of spin sublattices at integer fillings $\nu=1$ and $2$ with intralattice and interlattice exchange coupling $J^{\parallel}$ and $J^{\perp}$. Right panel: Temperature dependence of $\Delta |g|$ (top) and difference of the respective magnetic susceptibility $\Delta\chi_S$ in MCMC simulations (bottom). \textbf{d}, Upper panel: Top-view schematic of the bilayer lattice for integer fillings $\nu=1$ and $2$, with antiferromagnetic spin-exchange between rigidly locked Mott insulating electrons in the MoSe$_2$ layer (with interactions among the first, second and third nearest neighbors $J_1^{\text{RKKY}}$, $J_2^{\text{RKKY}}$ and $J_3^{\text{RKKY}}$) and mobile electrons in the WS$_2$ layer. Central panel: Results for nearest-neighbor RKKY interactions in the MoSe$_2$ layer at filling fractions $\nu > 1$ from simulations of non-interacting fermions (light blue) and TPSC (dark blue). Lower panel: Longer-range RKKY interactions can act against (or in favor of) the $120^{\circ}$ state depending on the filling factor as highlighted by the grey (white) areas. Close to the Mott insulating regime $\nu=2$ (hatched area), the perturbative approach breaks down and other effects such as kinetic magnetism are likely to dominate.}
\label{fig3}
\end{figure*}

In the simulation, the CB offset between the first DOS peaks in MoSe$_2$ and WS$_2$ is fixed to $30$~meV by the field $F_0$, with an uncertainty of about $10\%$ stemming from uncertainties in the thickness and the dielectric susceptibility of the HBL. The small CB offset confirms the near-resonant alignment in MoSe$_2$/WS$_2$~\cite{Zhang2016} and allows to tune the effective energetic ordering of the two layers by out-of-plane electric fields. Importantly, to explain the non-uniform width of region~I, the Coulomb gap $U$ between the first and second electron charging events in MoSe$_2$ must be larger than the CB offset. This is at the origin of the difference between parallel and antiparallel MoSe$_2$/WS$_2$ HBLs, and for the latter, our calculation predicts $U\approx 60$~meV which leads to the peculiar charging behavior illustrated in Fig.~\ref{fig2}e and g. At zero field, in a first step, all electrons fill the moir\'e potential minima inside MoSe$_2$ until the filling factor of $\nu = n/n_0=1$ is reached. In a second step, successive electrons occupy the WS$_2$ layer while the charge density on the MoSe$_2$ sublattice remains constant. Only after both layers host one electron per moir\'e site does the charge carrier doping into the primary MoSe$_2$ lattice continue. For $F \ll 0$~V/nm, on the contrary, the effective CB offset becomes larger than $U$, such that the second electron populates the MoSe$_2$ layer, leading to the continuous transition from region~I to region~II (cp. the left panel of Fig.~\ref{fig2}b).

We emphasize that while the electrons fill the WS$_2$ lattice, the optical response of both $M_1^-$ and $\tilde{M}_2^-$ is only marginally affected. This indicates that the moir\'e potential minima in WS$_2$ are located away from both the XX and the MM sites, and implies that the two excitons act as remote sensors of the emerging secondary lattice. To confirm this scenario, we consider $\tilde{M}_2^-$ pinned on the MM site of the moir\'e unit cell and subjected to Coulomb interactions with electron lattices of varying geometry for different fractional fillings. Using the variational approach (see Methods for details), we calculate the change in the exciton binding energy in the presence of the two laterally and vertically displaced charge lattices in MoSe$_2$ and WS$_2$ in the process of filling. The quantitative agreement between experiment and theory in Fig.~\ref{fig2}f is compelling: as the filling factor is increased from zero to two electrons per moir\'e cell, the binding energy varies from zero up to a maximum of $2$~meV, providing an estimate for the energy scale of interactions between excitons and electrons ordered on the surrounding vertically offset and laterally staggered moir\'e lattices with schematics in Fig.~\ref{fig2}g.


The presence of stabilized electron order in MoSe$_2$ between $\nu=1$ and $2$ enables detailed studies of the corresponding spin-lattice, with moir\'e excitons as local probes of magnetization. Previously, isolated bands in TMD moir\'e structures have been theoretically predicted~\cite{MacdonaldHubbard2018} and experimentally observed~\cite{Tang2020,Gerardot_exciton-polarons_2022} to mimic the triangular-lattice Hubbard model, which maps onto the spin Heisenberg model with antiferromagnetic order for strong on-site Coulomb repulsion $U$ and next-neighbor coupling only~\cite{MacdonaldHubbard2018}. Experimentally, spin-spin interactions manifest in diverging magnetic susceptibility which can be probed either by magnetically induced circular dichroism (MCD)~\cite{Ciorciaro2023May} or renormalized exciton $g$-factors~\cite{Tang2020,Gerardot_exciton-polarons_2022}. The former approach has been successfully adapted to probe kinetic magnetism in R-type MoSe$_2$/WS$_2$ HBLs~\cite{Ciorciaro2023May}, whereas we use the latter method and focus on the effective $g$-factor of $\tilde{M}_2^-$ as the main probe of spin polarization. The corresponding probe of magnetism by MCD of $M_1^-$ is shown in the Extended Data Fig.~\ref{SMfigMCD}.

Figure~\ref{fig3}a shows the temperature-dependent Zeeman splitting $\Delta E_Z = E^{+}-E^{-}$ between the $\tilde{M}_2^-$ peaks with $\sigma^+$ and $\sigma^-$ polarization for a range of discrete temperatures between $0.1$ and $28$~K at the filling factor $\nu=1$ in sample S2. Analogous to hole-mediated magnetism in WS$_2$/WSe$_2$~\cite{Tang2020} or MoSe$_2$/WSe$_2$~\cite{Gerardot_exciton-polarons_2022}, the evolution of $\Delta E_Z$ with magnetic field is highly nonlinear, with maximum slopes proportional to the effective $g$-factor with absolute values $|g|$ exceeding $600$ at the lowest temperature. A Curie-Weiss fit to the temperature-dependence of $|g|$ in the bottom left inset of Fig.~\ref{fig3}a confirms the paramagnetic response of the underlying spin lattice. Notably, the fit yields a negative Curie temperature $\theta = -40 \pm 15$~mK, which is consistent with weak antiferromagnetic interactions. The top right inset of Fig.~\ref{fig3}a shows the Zeeman splitting as a function of the rescaled field $B/T$, and the scaling collapse confirms that the degree of spin-polarization is limited by the thermal energy scale only.

The evolution of the $g$-factor upon electron doping within the stability range of $\tilde{M}_2^-$ (i.e. between $\nu=0$ and $2.5$) is shown in the upper panel of Fig.~\ref{fig3}b. For all temperatures, the absolute $g$-factor values rise quickly around the filling factor of $0.5$ until they reach their maximum at $\nu=1$. Remarkably, and in contrast to other heterostructures \cite{Tang2020,Gerardot_exciton-polarons_2022}, $|g|$ remains at high values as long as the peak $\tilde{M}_2^-$ is present. This is consistent with the understanding developed above: the MoSe$_2$ electron sublattice is locked in a Mott insulating state during successive charging of the WS$_2$ layer ($1<\nu <2$). Throughout the plateau, the $g$-factor values exhibit variations on the order of $10 - 15\%$ due to emergent filling of the secondary lattice. 
The corresponding variations are also pronounced in the Curie temperature $\theta$, determined from the respective Curie-Weiss fits to the data as a function of the filling factor and shown in the lower panel of Fig.~\ref{fig3}b. Akin to $g$-factors, $\theta$ exhibits variations across the range $1< \nu <2$ with two distinct maxima around $\nu=1$ and $\sim 1.8$. Since negative Curie temperatures are associated with antiferromagnetic (AFM) ordering, an increase in $\theta$ towards zero indicates a weakening of AFM interactions within doping regimes highlighted in grey in the bottom panel of Fig.~\ref{fig3}b.


To understand this behavior, we first focus on the $g$-factors at integer filling and plot the difference $\Delta |g| = |g|(\nu = 1)-|g|(\nu = 2)$ in the top panel of Fig.~\ref{fig3}c as a function of the temperature. Obviously, the $g$-factor of the bilayer lattice ($\nu = 2$) is always below the value of the singular lattice ($\nu = 1$) throughout the experimental temperature range. This can be understood qualitatively by invoking classical Heisenberg models on the corresponding mono- and bilayer lattices and computing the magnetic susceptibility $\chi_S$ by Markov chain Monte Carlo methods (MCMC) (see Methods for details). The difference in magnetization 
 $\Delta\chi_S=\chi_S(\nu = 1)-\chi_S(\nu = 2)$ computed for $J^\perp /J ^\parallel=1$, shown in the bottom top panel of Fig.~\ref{fig3}c, confirms the experimentally observed trend. This aligns with the intuition of the magnetic structure in the bilayer Hubbard model: coupling two singular triangular lattices in antiferromagnetically ordered $120^{\circ}$ state by finite interlayer coupling $J^{\perp}$ results in enhanced frustration between the magnetic moments, canting the lattice-ordered spins out-of-plane. This in turn suppresses correlations within each layer, which results in a reduction of the susceptibility as compared to the case of a singular filled spin lattice. This qualitative correspondence between the classical spin model and the observed phenomenology of $g$-factors at integer fillings $\nu=1$ and $2$ suggests a non-vanishing antiferromagnetic interlayer coupling $J^{\perp}$ and supports the view that the system is governed by bilayer Fermi-Hubbard physics~\cite{Xu2022Sep}.

Now we examine the regime $1< \nu < 2$, where the primary MoSe$_2$ layer is locked in a Mott insulating state while the WS$_2$ layer fills up. For low doping of the secondary lattice, the system can be described by a weakly interacting Fermi liquid coupled to local moments in a Mott insulator. As illustrated in the schematics of Fig.~\ref{fig3}c and d, finite interlayer coupling $J^{\perp}$ introduces interactions between mobile electrons on the WS$_2$ sublattice and localized spins in the MoSe$_2$ lattice, leading to emergent RKKY-type intralayer interactions $J^{\text{RKKY}}$~\cite{Ruderman1954,Coleman_2015}. We approximate the strength of RKKY interactions by assuming free fermions in the WS$_2$ layer and employing the two-particle self-consistent (TPSC) approach~\cite{VT1997,Tremblay2012} (see Methods). The central panel of Fig.~\ref{fig3}d shows the corresponding ferromagnetic RKKY corrections $J^{\text{RKKY}}_{1}$ to the nearest-neighbor intralayer interactions $J^{\parallel}_1$ (with effective nearest-neighbor intralayer interactions on the MoSe$_2$ sublattice $J^{\parallel}_1 + J^{\text{RKKY}}_1 < J^{\parallel}_1$). For low doping above unity filling of the primary sublattice ($\nu \gtrsim 1$), a rapid drop of $J^{\text{RKKY}}_1$ is observed (central panel of Fig.~\ref{fig3}d), which tends to increase the Curie temperature around $\nu=1$ (bottom panel of Fig.~\ref{fig3}b). This is also consistent with the initial drop of the $g$-factor value for $\nu \gtrsim 1$: reduced interactions increase the effective spin temperature in the MoSe$_2$ sublattice, which results in reduced susceptibility.  

As doping proceeds ($\nu \gtrsim 1.5$), nearest-neighbor interactions are further renormalized, resulting in a second rise in the Curie temperature towards zero in Fig.~\ref{fig3}b. The results of the free fermion and TPSC calculations in the central panel of Fig.~\ref{fig3}d indicate that interactions in the WS$_2$ layer enhance this effect. In experiments between $1< \nu \lesssim 1.8$, the Curie temperature exhibits a minimum at $\nu \approx 1.4$, not anticipated from simple nearest-neighbor RKKY interactions. A possible origin for this behavior is provided by longer-range RKKY couplings, shown in the lower panel of Fig.~\ref{fig3}d for straight second- and third-nearest neighbors interactions $J^{\text{RKKY}}_2$ and $J^{\text{RKKY}}_3$. Indeed, these higher-order effects show distinct oscillations that can act in favor or against the $120^{\circ}$ AFM order: positive (negative) second- (third-) neighbor interactions act in favor of the $120^{\circ}$ state and imply a decrease in the Curie temperature $\theta$, whereas a reversed sign inhibits the $120^{\circ}$ order as in doping regions highlighted by grey shaded areas in Fig.~\ref{fig3}b and d. Away from these doping regimes (as for $1.15 \lesssim \nu \lesssim 1.45$), AFM order is supported, as signified by reduced Curie temperatures in Fig.~\ref{fig3}b. 

Close to the regime $\nu \lesssim 2$ (grey hatched area in Fig.~\ref{fig3}d), the Fermi liquid picture of the WS$_2$a layer breaks down and a bilayer Mott insulator forms, implying that other effects of strongly-correlated origin likely take over. In particular, slight underdoping of a Mott insulator on a triangular lattice is known to result in kinetic effects of Haerter-Shastry type antiferromagnetism, which strongly favors the formation of a classical $120^{\circ}$ state to minimize the kinetic energy of the vacancies~\cite{Haerter2005,Sposetti2014,Morera2023Jun,Schlömer2023kinetic}. Such kinetic effects could explain the rise in $\theta$ above doping $\nu \approx 1.8$ until $\nu=2$ in Fig.~\ref{fig3}b, overcompensating the effect of RKKY interactions. This would align with the behavior of $g$-factors at $\nu= 2-\epsilon$,
close to the bilayer Mott insulator, where kinetic magnetism is expected to support AFM correlations and thus enhance the susceptibility. Kinetic magnetism in the regime at $\nu= 1-\epsilon$ could be also responsible for the sub-plateau on the rising flank of the $g$-factor values in Fig.~\ref{fig3}b for fillings between $0.7$ and $0.9$ at $T=0.16$~K, which vanishes at higher temperatures. First, around $\nu = 0.75$, the electrons in the MoSe$_2$ layer become increasingly frustrated, leading to an initial saturation of the $g$-factor. However, for $\nu \gtrsim 0.9$, the release of kinetic frustration via Haerter-Shastry type AFM interactions would promote the $g$-factors to significantly higher values.  


We conclude by pointing out that even though many aspects of magnetism on the bilayer triangular spin-charge lattice of antiparallel MoSe$_2$/WS$_2$ heterostructures with pronounced antiferromagnetic corrrelations are conclusively supported by our theory, generalized Wigner crystal formation with additional electron-ordering effects away from integer-filling Mott insulating states \cite{Xu2020} may lead to renormalizations of magnetic interactions at certain doping levels. Complementary to parallel MoSe$_2$/WS$_2$ heterostacks~\cite{Ciorciaro2023May}, our results establish antiparallel stacking as an even richer platform for manifestations of kinetic magnetism and other theoretically predicted phases in related settings~\cite{MacdonaldHubbard2018,Morales-Duran2022May,Hu2021Dec,Zang2021Aug,Devakul2021Nov,Morera2023Jun,Tao2023Jul,Seifert2024Jan}, providing compelling motivation for future experimental and theoretical work on many-body phenomena and magnetism in MoSe$_2$/WS$_2$ moir\'e bilayer Hubbard lattices.

\section*{Methods}
\noindent \textbf{Device fabrication:} Monolayers of MoSe$_2$ and WS$_2$ were either exfoliated from bulk crystals (HQ Graphene) or obtained from chemical vapor deposition. Thin flakes of hBN were exfoliated from bulk crystals (NIMS). Devices from hBN-encapsulated MoSe$_2$/WS$_2$ HBLs were prepared by standard dry exfoliation-transfer method. More details on samples are given in the Supplementary Information. 
\vspace{8pt}
\\    
\noindent \textbf{Optical spectroscopy:} Cryogenic DR spectroscopy was conducted using home-built confocal microscopes in back-scattering geometry. The samples were loaded into a closed-cycle cryostat (attocube systems, attoDRY1000) with a base temperature of $3.2$~K or a dilution refrigerator (Leiden Cryogenics) operated between $0.1$ and $26$~K. Both cryogenic systems were equipped with a superconducting magnet providing magnetic fields of up to $\pm 9$~T in Faraday configuration. Piezo-stepping and scanning units (attocube systems, ANPxyz and ANSxy100) were used for sample positioning with respect to a low-temperature apochromatic objective (attocube systems, LT-APO/VISIR0.82 and NIR0.81). For DR measurements, a stabilized Tungsten-Halogen lamp (Thorlabs, SLS201L) and supercontinuum lasers (NKT Photonics, SuperK Extreme and SuperK Varia) were used as broadband light sources. The reflection signal was spectrally dispersed by monochromators (Roper Scientific, Acton SP2500 or Acton SpectraPro 300i with 300 grooves/mm gratings) and detected by liquid nitrogen or Peltier cooled charge-coupled devices (Roper Scientific, Spec-10:100BR or Andor, iDus 416). A set of linear polarizers (Thorlabs, LPVIS), half- and quarter-waveplates (B. Halle, $310-1100$~nm achromatic) mounted on piezo-rotators (attocube systems, ANR240) was used to control the polarization in excitation and detection. The DR spectra were obtained by normalizing the reflected spectra from the HBL region ($R$) to that from the sample region without MoSe$_2$ and WS$_2$ layers ($R_0$) as $\textrm{DR} = (R-R_0)/R_0$. 
\vspace{8pt}
\\
\noindent\textbf{Electrostatic simulations:}
We follow the approach outlined in Refs.~\cite{Tan2023May, Popert2022, Ma2021Oct,Mak_Shan_WSe2_iX_2018}. We denote the chemical potential inside the top (bottom) layer by $E_{T (B)}$, emphasizing that it acts as an energy cost per charge carrier induced into the monolayer. The charge density in a monolayer is given by the integrated DOS, $n_{i} (E_i) \, = \, \int_{0}^{E_i} \, \text{DOS}_i(E') dE'$, with $i=T , B$. We note that at this level of description, the DOS is explicitly not related to a single particle band structure, since electrons might also exhibit energy costs from mutual Coulomb repulsion. The electrostatic equations are then given by: 
\begin{equation}
e\begin{pmatrix}
n_T  \\
n_B 
\end{pmatrix} = 
\begin{pmatrix}
C_g^T & -C_g^T - C_S & C_S & 0 \\
0 & C_S & -C_g^B - C_S & C_g^B
\end{pmatrix}
\begin{pmatrix}
V_{TG} \\
E_T/e \\
E_B/e \\
V_{BG}
\end{pmatrix}\, ,
\end{equation}
with $C_g^{T(B)} = \epsilon_0 \epsilon_\text{\tiny{hBN}} / d_{T(B)}$ being the geometric and $C_S = \epsilon_0 \epsilon_\text{\tiny{TMD}} / d_S$ the HBL capacitance. Here, we use $\epsilon_\text{\tiny{hBN}}  = 4$, $\epsilon_\text{\tiny{TMD}}=8$, $d_{T(B)} = 55$~nm (as determined with atomic force microscopy), and $d_{S} = 0.6$~nm. For a given $\text{DOS}_i$, we solve these equations numerically for $(E_T, E_B)$ and subsequently evaluate $n_{T(B)}$ at the computed energies. Finally, we assume that charging occurs in steps of $n_0$, with each step corresponding to a peak in $\text{DOS}_i$, and fit the charging behavior in Fig.~\ref{fig2}a by adjusting the energies of the peaks in $\text{DOS}_i$ as shown in Fig.~\ref{fig2}d.
\vspace{8pt}
\\
\noindent \textbf{Coulomb-interaction energy:}
To calculate the interaction energy (binding energy) of moir\'e excitons with electrons ordered on two vertically displaced and staggered lattices, we assume that the exciton is confined in one moir\'e cell and interacts with the surrounding ordered electrons as illustrated in Fig.~\ref{fig2}g. 
We further assume that the main contribution to the binding energy stems from charge-induced modification of the electron-hole relative motion $\bm \rho \equiv (\rho,\theta) = \mathbf{r}_\text{e} - \mathbf{r}_\text{h}$, where $\mathbf{r}_\text{e(h)}$ are the coordinates of the electron and hole forming the exciton. The corresponding Schr\"odinger equation takes the form:
\begin{equation}
  -\frac{\hbar^2}{2\mu} \Delta \varphi(\bm\rho) + [V_\text{RK}(\rho) + V(\bm\rho)] \varphi(\bm\rho) = E \varphi(\bm\rho), \nonumber
\end{equation}
where $E$ is the exciton energy, $\mu = m_\text{e} m_\text{h}/(m_\text{e}+m_\text{h})$ is the reduced exciton mass, $m_\text{e}$ and $m_\text{h}$ are the electron and hole effective masses, and the Rytova--Keldysh potential~\cite{Rytova1967,Keldysh1979} of the electron-hole attraction is given by:
\begin{equation}
  V_\text{RK}(\rho) = -\frac{\pi e^2}{2 \varepsilon \rho_0} \left[ H_0 \left(\frac{\rho}{\rho_0}\right) - Y_0 \left(\frac{\rho}{\rho_0}\right) \right]. \nonumber
\end{equation}
Here, $e$ is the electron charge, $\rho_0$ is the screening length, $\varepsilon$ is the effective dielectric constant, and $H_0 (x)$ and $Y_0 (x)$ are the Struve and Neumann functions. 

The interaction of the exciton with the charge lattice is described by the Coulomb sum:
\begin{equation}
V(\bm\rho) = \pm \frac{e^2}{\varepsilon} \sum_\mathbf{n}
                                \left[
                                  \frac{1}{|\beta_\text{e} \bm\rho + \mathbf{n}|}
                                  -\frac{1}{|\beta_\text{h} \bm\rho - \mathbf{n}|}
                                \right],
                                \nonumber
\end{equation}
where the plus and minus signs correspond to positive and negative elementary charges, $\beta_\text{e} = m_\text{e}/(m_\text{e}+m_\text{h})$, $\beta_\text{h} = m_\text{h}/(m_\text{e}+m_\text{h})$, and $\mathbf{n}$ are the coordinates of electrons/holes on the lattice. The two terms in the brackets determine the interaction of the charge lattice with the hole and the electron that constitute the exciton. 

To determine the binding energy of the state, we calculate the free exciton energy $E_X$ to obtain:
\begin{equation}
  E_{b} = E_X - E. \nonumber
\end{equation}
To calculate $E_X$, we set $V(\bm \rho)=0$, and use in the calculations of both $E_X$ and $E$ the set of 2D hydrogen-like wave functions with the Bohr radius as variational parameter~\cite{Chernikov2014,Courtade2017,Semina2019} and the basis of six functions~\cite{Yang1991} with quantum numbers $(n,l) = (1,0),(2,0),(2,\pm1),(4,\pm3)$ to take into account polarization effects on the exciton relative motion. Due to the lower rotational symmetry of the potential $V(\bm\rho)$, we also include hydrogen-like wave functions with angular momenta $l = \pm 1, \pm 3$. The explicit expression for the trial function is:
\begin{equation}
  \varphi(\rho,\theta) = e^{-\alpha \rho}
                          + \zeta \rho e^{-\beta \rho}
                           + \eta \rho e^{-\gamma \rho} \cos\theta
                            + \xi \rho^3 e^{-\delta \rho} \cos3\theta.
                             \nonumber
\end{equation}
We solve the minimization problem numerically for seven parameters $(\alpha, \beta, \gamma, \delta, \zeta, \eta, \xi)$ using MATLAB R2017B and experimental material parameters of MoSe$_2$ monolayers~\cite{Goryca2019}: $m_\text{e} = 0.84 m_0$, $m_\text{h} = 0.6m_0$, $\varepsilon = 4.4$, $\rho_0 = 0.89$~nm. The only fitting parameter for comparison between the experimental data and the theoretical model is the moir\'e superlattice constant, which in Fig.~\ref{fig2}f is taken to be $7.7$~nm.

\vspace{8pt}
\noindent \textbf{Markov chain Monte Carlo simulations:}
In the Mott insulating regimes of $\nu = 1$ and $2$, we model the system by a classical Heisenberg spin model on the singular and bilayer triangular lattice, respectively, with the Hamiltonian
\begin{equation}
H = J \sum_{\braket{\mathbf{i},\mathbf{j}}} \mathbf{S}_{\mathbf{i}} \cdot \mathbf{S}_{\mathbf{j}}.
\end{equation} 
Here, $\mathbf{S}_{\mathbf{i}}$ are classical vectors of length $|\mathbf{S}_{\mathbf{i}}| = 1$ at lattice site $\mathbf{i}$. Using Markov chain Monte Carlo methods, we compute the susceptibility given by 
\begin{equation}
\chi_s = \beta V (\braket{m^2} - \braket{m}^2),
\end{equation}
where $\beta$ is the inverse temperature, $V$ is the number of lattice sites on a grid of $V = L_x \times L_y$ lattice sites ($32 \times 32$ in our simulations) with periodic boundary conditions, and $m$ is the absolute value of the magnetization averaged over a snapshot:
\begin{equation}
m = \sqrt{\sum_{\mu = x,y,z} m_{\mu}^2}, \quad m_{\mu} = \frac{1}{V} \sum_{\mathbf{i}} S^{\mu}_{\mathbf{i}}.
\end{equation} 
The simulation results are shown as the difference in the magnetic susceptibility for filling factors $\nu = 1$ and $2$ in Fig.\ref{fig3}c.

\vspace{8pt}
\noindent \textbf{Two-particle self-consistent theory:}
We calculate the spin susceptibility of the Hubbard model on the triangular lattice using the two-particle self-consistent (TPSC) approach introduced by Vilk and Tremblay~\cite{VT1997, Tremblay2012}. Here, the central idea is to renormalize the interaction vertex away from the bare Hubbard interaction $U$ for a given filling within the random phase approximation (RPA), such that the (spin) susceptibility reads
\begin{equation}
    \chi_{\text{sp}}(q) = \frac{\chi_0(q)}{1 - \frac{1}{2} U_{\text{sp}} \chi_0(q) },
\end{equation}
where $\chi_0(q) = \chi_0(\mathbf{q}, i\omega_{\nu})$ is the non-interacting polarization bubble given by 
\begin{equation}
\chi_0(q) = -2 \frac{T}{N_{\mathbf{k}}} \sum_{\mathbf{k}} \frac{n_F(\epsilon_{\mathbf{k}} - \mu) - n_F(\epsilon_{\mathbf{k} + \mathbf{q}} - \mu)}{i\omega_n + \epsilon_{\mathbf{k}} - \epsilon_{\mathbf{k} + \mathbf{q}}},
\end{equation}
with the single-particle dispersion $\epsilon_{\mathbf{k}}$, the chemical potential $\mu$, the Fermi-Dirac distribution $n_F(x)$, the number of sites $N_{\mathbf{k}}$, and the bosonic Matsubara frequencies $i\omega_{\nu}$. In our case, $\epsilon_{\mathbf{k}} = -2t [\cos (k_x) + \cos (k_y) + \cos (k_x + k_y)]$, i.e. we map the triangular lattice onto a square lattice with next-nearest neighbor diagonal hopping terms featuring identical tight-binding physics. 

The following local sum rule for the spin susceptibility can be derived from the Bethe-Salpeter equation by enforcing Pauli's principle:
\begin{equation}
    \begin{aligned}
     \frac{T}{N_{\mathbf{q}}} \sum_{\mathbf{q}, i\omega_n} \chi_{\text{sp}}(\mathbf{q}, i\omega_n) = n  - 2 \braket{\hat{n}_{\uparrow} \hat{n}_{\downarrow}},
    \end{aligned}
    \label{eq:TPSC}
\end{equation}
where $\hat{n}_{\sigma}$ is the particle density of spin species $\sigma$, and spin balance $\braket{\hat{n}_{\uparrow}} = \braket{\hat{n}_{\downarrow}} = n/2$ is assumed. Vilk and Tremblay introduced the following ansatz between $\braket{\hat{n}_{\uparrow} \hat{n}_{\downarrow}}$ and $U_{\text{sp}}$:
\begin{equation}
    \frac{U_{\text{sp}}}{U} = \frac{\braket{\hat{n}_{\uparrow} \hat{n}_{\downarrow}}}{n_{\sigma}^2},
    \label{eq:ansatz}
\end{equation}
which, together with Eq.~\eqref{eq:TPSC}, defines the two-particle self-consistent theory. The results of the respective simulations are shown in Fig.~\ref{fig3}d. 

\vspace{8pt}
\noindent \textbf{RKKY interactions:}
In the regime $\nu \gtrsim 1$, the MoSe$_2$ layer is in a Mott insulating state, while the WS$_2$ layer is assumed to be metallic (i.e. void of generalized Wigner crystallization). Due to finite interlayer Kondo-type electronic coupling $J^{\perp}$, a magnetic moment in the MoSe$_2$ layer at site $\mathbf{i}$ induces a magnetization profile in the metal, which in turn couples to another localized spin at position $\mathbf{j}$. This results in the emergence of long-range interactions between the localized spins, described by the Hamiltonian (the factor of $1/2$ prevents double counting)~\cite{Ruderman1954,Coleman_2015}:
\begin{equation}
 \hat{\mathcal{H}}^{\text{RKKY}} = \frac{1}{2} \sum_{\mathbf{i}, \mathbf{j}} J^{\text{RKKY}}_{\textbf{i} \textbf{j}} \hat{\mathbf{S}}_{\mathbf{i}} \cdot \hat{\mathbf{S}}_{\mathbf{j}}.
\end{equation}
The interaction $J^{\text{RKKY}}_{\textbf{i} \textbf{j}}$ is calculated from the susceptibility of the metal. As each site in the MoSe$_2$ layer interact with three sites in the WS$_2$ layer (assuming nearest-neighbor interlayer couplings), the contributions to $J^{\text{RKKY}}_{\textbf{i} \textbf{j}}$ read (see Extended Data Fig.~\ref{fig:RKKY_fig})~\cite{Ruderman1954,Coleman_2015}:
\begin{equation}
J^{\text{RKKY}}_{\textbf{i} \textbf{j}} = -(J^{\perp})^2 \sum_{\mathbf{a} \in \{ 1,2,3 \}} \sum_{\mathbf{b} \in \{ 4,5,6\}} \chi(\mathbf{a} - \mathbf{b}),
\label{eq:RKKY_int}
\end{equation}  
with
\begin{equation}
\chi(\mathbf{x}) = \int_{\mathbf{q}} d\mathbf{q} e^{i\mathbf{q}\cdot \mathbf{x}} \chi(i\omega_n = 0, \mathbf{q}).
\end{equation}
The simulation results for $J^{\text{RKKY}}_{1}$, $J^{\text{RKKY}}_{2}$, and $J^{\text{RKKY}}_{3}$ are shown in Fig.\ref{fig3}d.

\vspace{8pt}
\noindent \textbf{Acknowledgements:}\\
We thank Ata\c{c} Imamo\u{g}lu and Eugene Demler for fruitful discussions. This research was funded by the European Research Council (ERC) under the Grant Agreement No.~772195 (A.\,H.) and No.~948141 (F.\,G.), as well as the Deutsche Forschungsgemeinschaft (DFG, German Research Foundation) within the Priority Programme SPP~2244 2DMP and the Germany's Excellence Strategy EXC-2111-390814868 (MCQST). B.\,P. acknowledges funding by IMPRS-QST. I.\,B. acknowledges support from the Alexander von Humboldt Foundation. X.\,H. and A.\,S.\,B. received funding from the European Union's Framework Programme for Research and Innovation Horizon 2020 (2014--2020) under the Marie Sk{\l}odowska-Curie Grant Agreement No.~754388 (LMUResearchFellows) and from LMUexcellent, funded by the Federal Ministry of Education and Research (BMBF) and the Free State of Bavaria under the Excellence Strategy of the German Federal Government and the L{\"a}nder. Z.\,L. was supported by the China Scholarship Council (CSC) No. 201808140196. A.\,H. acknowledges funding by the Bavarian Hightech Agenda within the EQAP project. K.\,W. and T.\,T. acknowledge support from the JSPS KAKENHI (Grant Numbers 20H00354 and 23H02052) and World Premier International Research Center Initiative (WPI), MEXT, Japan.
\vspace{8pt}
\\
\noindent \textbf{Contributions:}\\
J.\,G., Z.\,L. and I.\,B. synthesized monolayer crystals. K.\,W. and T.\,T. provided high-quality hBN. X.\,H., J.\,S., C.\,M. and S.\,M. fabricated field-effect devices. B.\,P., J.\,S., J.\,F., S.\,M. and J.\,T. performed optical spectroscopy. B.\,P. developed and performed electrostatic simulations. A.\,S.\,B.  developed and performed numerical calculations on bound exciton states in periodic charge lattices. H.\,S., A.\,B. and F.\,G. developed and performed MCMC Heisenberg model simulations and numerical TPSC calculations for correlated many-body systems. B.\,P., J.\,S., S.\,M. and A.\,H. analyzed the data. B.\,P., H.\,S., A.\,S.\,B. and A.\,H. wrote the manuscript. All authors commented on the manuscript. B.\,P., J.\,S., and S.\,M. contributed equally to this work.
\vspace{8pt}
\\
\noindent \textbf{Corresponding authors:}\\
B.\,P. (borislav.polovnikov@physik.uni-muenchen.de), H.\,S. (h.schloemer@physik.uni-muenchen.de) A.\,S.\,B. (anvar.baimuratov@lmu.de) and A.\,H. (alexander.hoegele@lmu.de). 
\vspace{8pt}
\\
\textbf{Data availability:} 
The data that support the findings of this study are available from the corresponding authors upon reasonable request.
\vspace{8pt}
\\
\textbf{Code availability:} 
The codes that support the findings of this study are available from the corresponding authors upon reasonable request.
\vspace{8pt}
\\
\textbf{Competing interests:}\\
The authors declare no competing interests.

%

\newpage

\renewcommand{\figurename}{Extended Data Fig.}
\setcounter{figure}{0}

\begin{figure*}[t!]  
\includegraphics[scale=1.0]{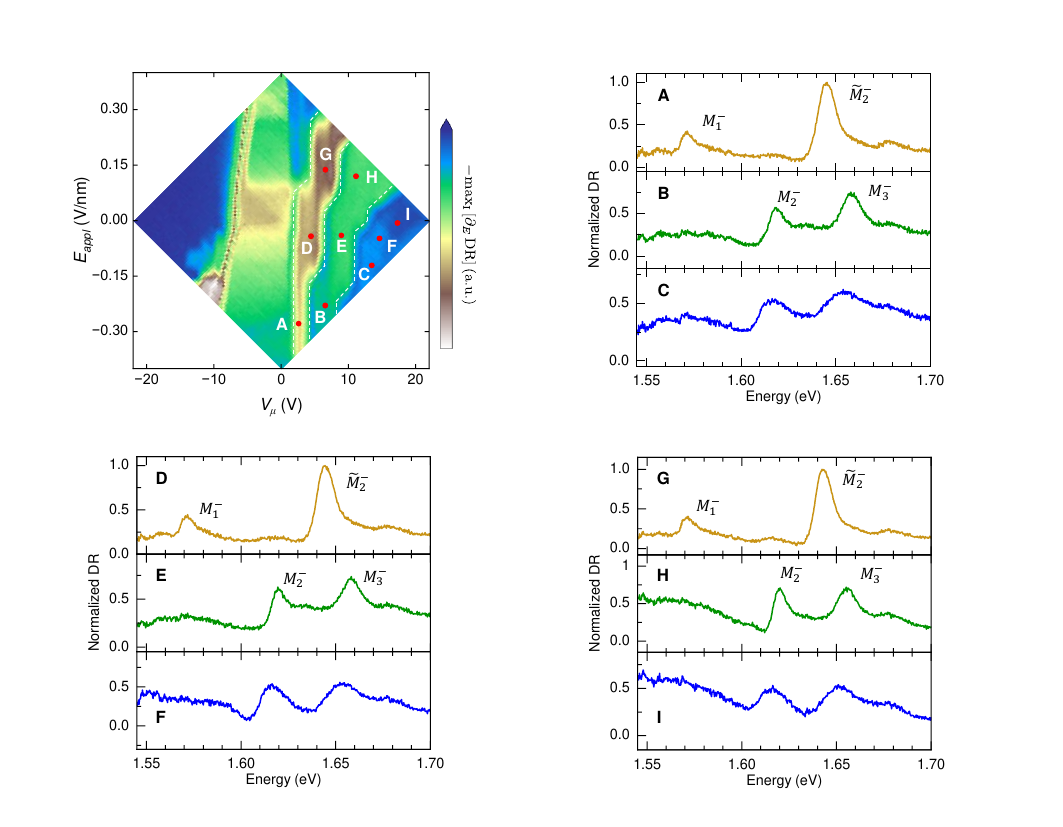}
\caption{\textbf{Evolution of DR in sample S1 with $\mathbf{V_{TG}}$ and $\mathbf{V_{BG}}$.} DR spectra at nine different points of the charging diagram (A-I) across the regions I, II and III. As obvious from direct comparison, the optical signatures of $M_1^-$, $\tilde{M}_2^-$, $M_2^-$, and $M_3^-$ remain at constant energies for voltages of same doping regimes.}
\label{SMfigSpectra}
\end{figure*}

\begin{figure*}[t!]  
\includegraphics[scale=1.0]{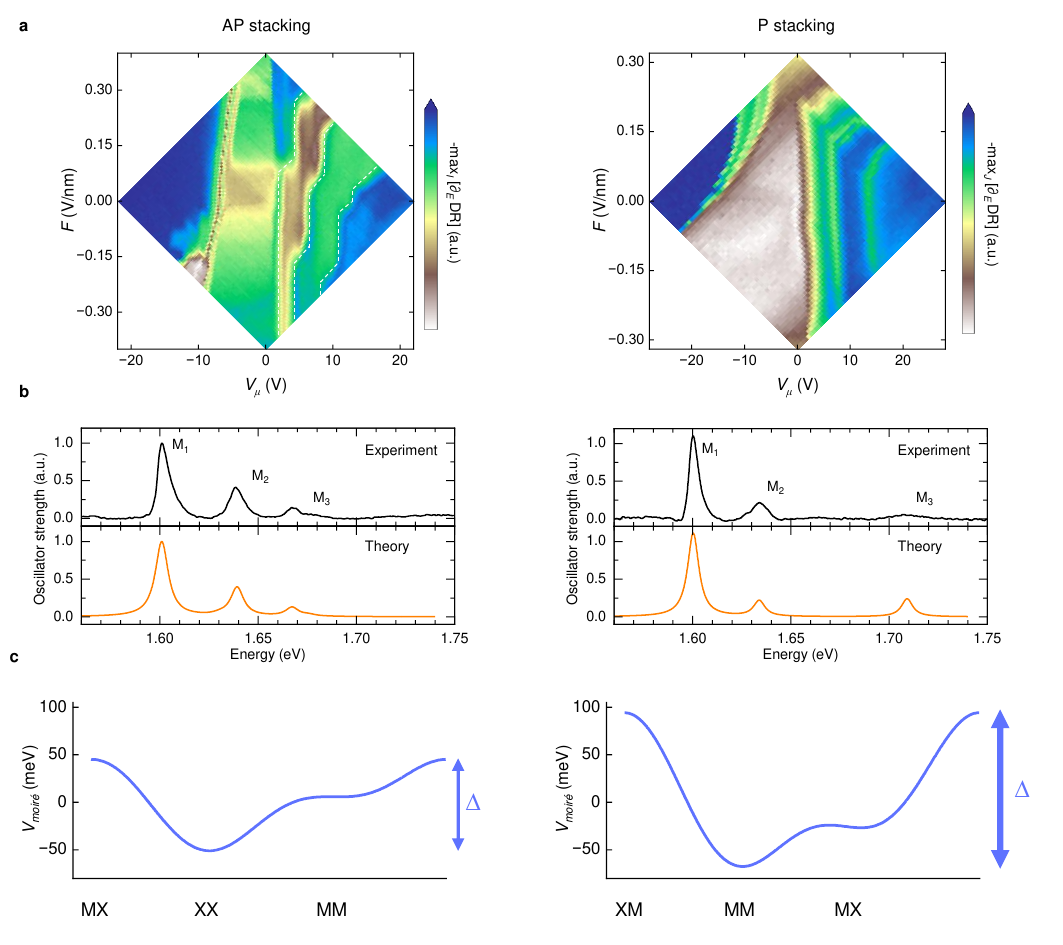}
\caption{\textbf{Comparison of antiparallel (AP) and parallel (P) MoSe$_{2}$/WS$_{2}$ charging behavior.} \textbf{a}, Hyper-spectral dual-gate DR data for antiparallel (left panel) and parallel (right panel) stackings, with the latter obtained from a sample assembled from monolayers synthesized by chemical vapor deposition. The interval $J$ for the visualization of data in the parallel heterostack is between $1.595$ and $1.800$~eV. Notably, the first charging step into the MoSe$_2$ layer does not change between negative and small positive electric fields in the parallel stack, signifying that all electrons charge the MoSe$_2$ layer. We attribute this behavior to a deeper moir\'e potential in parallel heterostacks, with a rough estimate obtained by fitting the neutral exciton spectra in \textbf{b} as detailed in Ref.~\cite{Polovnikov2023Apr}, with peak-to-peak potential amplitudes of $100$ and $170$~meV shown in \textbf{c} for the antiparallel (left panel) and parallel heterostack (right panel).}
\label{SMfigRType}
\end{figure*}

\begin{figure*}[t!]  
\includegraphics[scale=1.0]{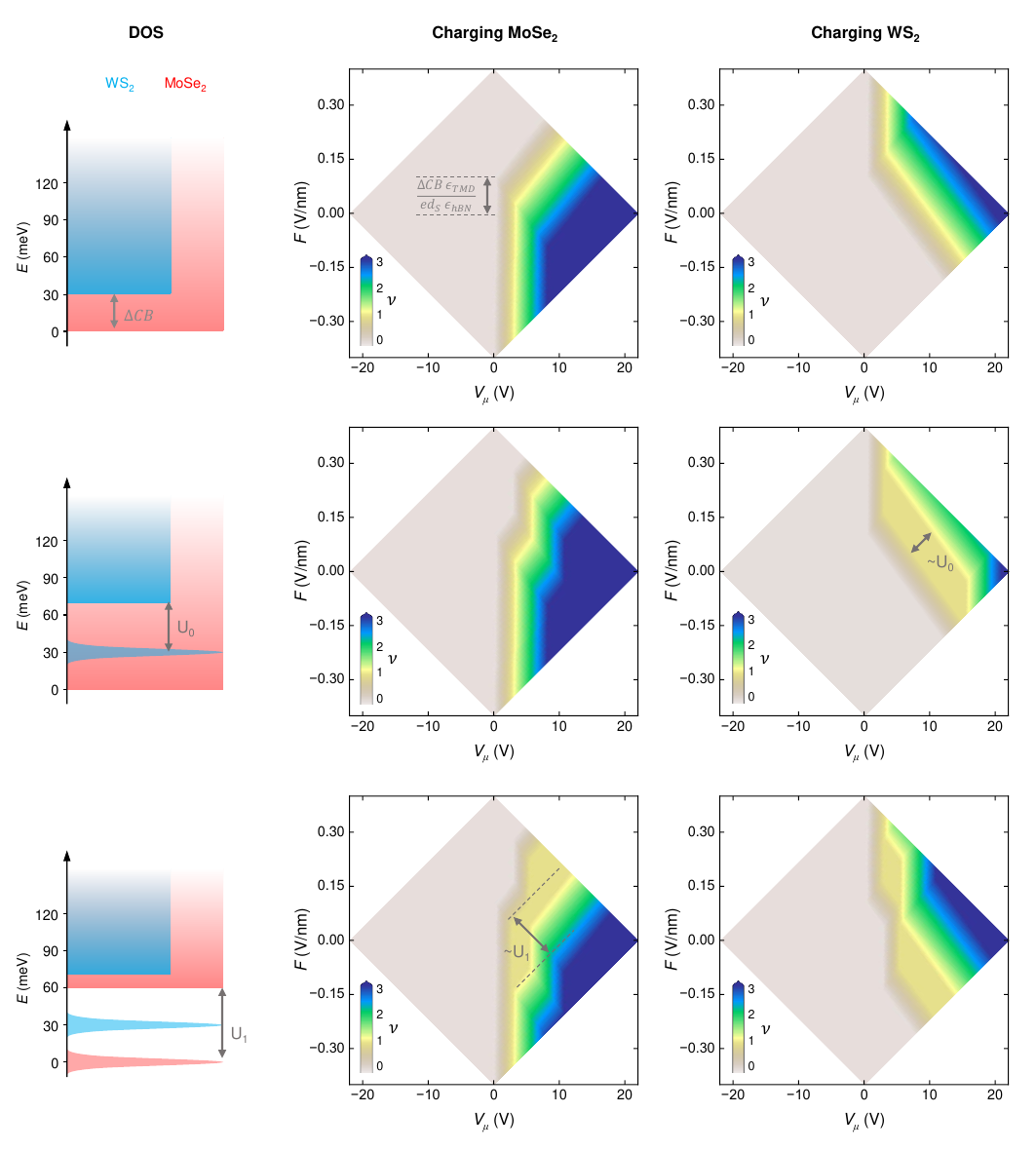}
\caption{\textbf{Electrostatic simulations with discretized density of states.} The computed charge distribution among the two layers (MoSe$_2$ and WS$_2$ in the central and right columns, respectively) shows distinct responses to different parameters in the simulations. The top panel highlights the effect of the conduction band offset $\Delta$CB for otherwise constant DOS (the higher DOS for MoSe$_2$ reflects its higher electron effective mass as compared to WS$_2$). The middle panel shows the effect of the energy gap in the DOS of WS$_2$, entailing step-like charging in WS$_2$ and kinks in the charge-onset line in MoSe$_2$. The bottom panel emphasizes the effect of gapped subbands in MoSe$_2$, resulting in step-like charging of the MoSe$_2$ layer.}
\label{SMfigModel}
\end{figure*}

\begin{figure*}[t!]  
\includegraphics[scale=1.0]{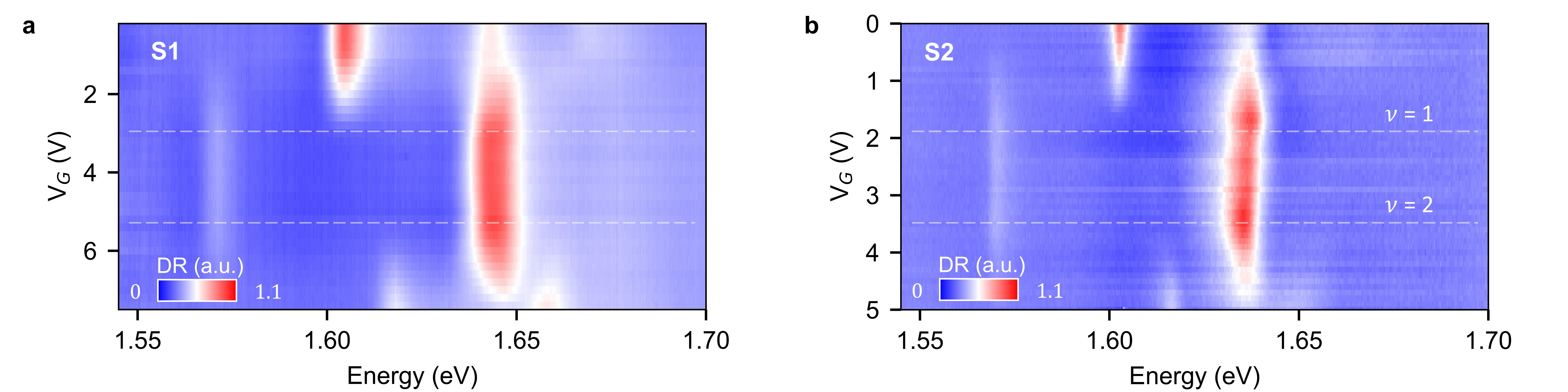}
\caption{\textbf{Direct comparison of charging in samples S1 and S2.} Evolution of normalized DR across the first charging step. Both samples feature the same charging behavior, with peaks $M_1^-$ and $\tilde{M}_2^-$ throughout extended ranges of voltages. Sample S2 (assembled from exfoliated native crystals) exhibits a narrower linewidth than sample S1 (assembled from monolayers synthesized by chemical vapor deposition).}
\vspace{2000pt}
\label{SMfigDevices}
\end{figure*}

\begin{figure*}[t!]  
\includegraphics[scale=1.0]{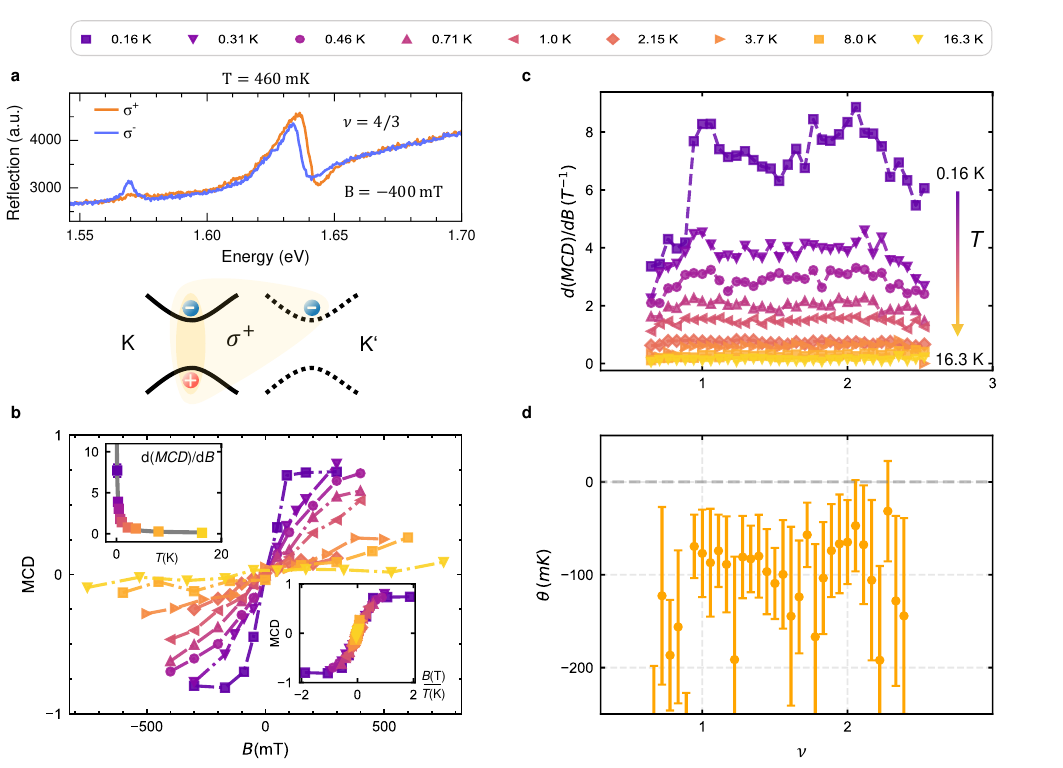}
\caption{\textbf{Magnetic circular dichroism (MCD) of $\mathbf{M_1^-}$ between $\mathbf{\nu =1}$ and $\mathbf{2}$ electrons per moir\'e cell.} \\ \textbf{a}, White light reflection spectra of the sample S2 in $\sigma^{+/-}$ polarization for $\nu = 4/3$, $B=-400$~mT and $T = 460$~mK. The $M_1^-$ peak at $1.57$~eV is strongly polarized, reminiscent of the canonical trion in monolayer MoSe$_2$ illustrated in the schematics. \textbf{b}, MCD of $M_1^-$, evaluated as the polarization contrast $\text{MCD} = (A_{\sigma^+} - A_{\sigma^-})/(A_{\sigma^+} + A_{\sigma^-})$, with $A_{\sigma^{+/-}}$ being the reflection contrast of the peak $M_1^-$ in $\sigma^+/-$ circular polarization, as a function of the magnetic field at $\nu=1$ \cite{Ciorciaro2023May}. As for the $g$-factors of $\tilde{M}_2^-$ shown in Fig.~\ref{fig3}, the slopes $d\text{(MCD)}/dB$ decrease with increasing temperature. The left inset shows the Curie-Weiss fit of the slopes at small fields, and the right inset shows the scaling collapse of the $\text{MCD}$ data as a function of $B/T$. \textbf{c}, The slopes of $\text{MCD}$ with respect to $B$ at small fields exhibit consistently high values throughout the stability regime of $M_1^-$, with fluctuations on the order of $10-15\%$ similar to the behavior of the $g$-factors of $\tilde{M}_2^-$. Across all temperatures there are two maxima around $\nu = 1$ and $2$ as well as a local minimum around $\nu=3/2$. \textbf{d}, Curie temperatures extracted from the MCD-slopes in \textbf{c} indicate weak antiferromagnetic exchange.}
\label{SMfigMCD}
\end{figure*}

\begin{figure*}[t!]  
\centering
\includegraphics[scale=1.0]{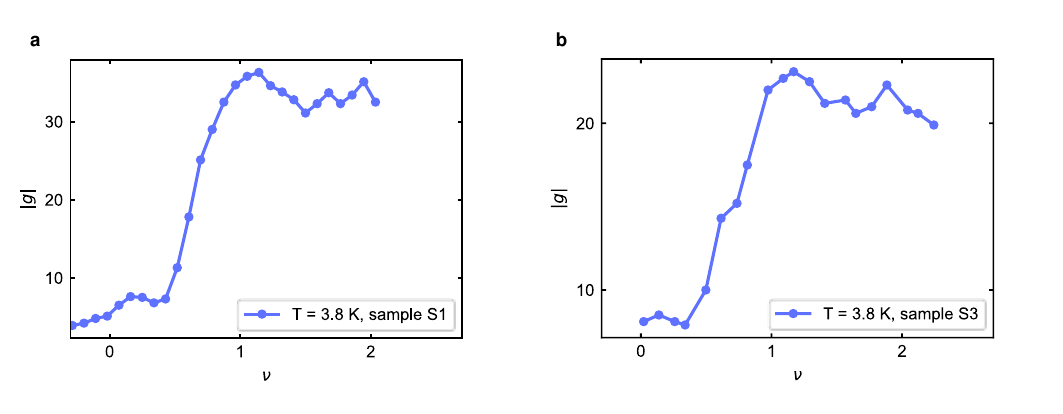}
\caption{\textbf{Experimental $g$-factors of $\tilde{M}_2^-$ as a function of the filling factor for two antiparallel MoSe$_2$/WS$_2$ heterostacks.} \textbf{a} and \textbf{b}, Absolute values of $g$-factors of $\tilde{M}_2^-$ in sample S1 and sample S3 (assembled from monolayers synthesized by chemical vapor deposition) at $3.8$~K, respectively, as a function of the electron filling factor. The evolution of the absolute values of $g$-factors with electron filling is similar for all samples S1, S2 and S3 with antiparallel alignment, exhibiting a rapid increase towards a plateau above $\nu = 1$ with moderate decrease and fluctuations upon further filling of the WS$_2$ layer.}
\vspace{1600pt}
\label{fig:SM_g_factors}
\end{figure*} 

\clearpage
\begin{figure*}[t!]  
\centering
\includegraphics[scale=1.0]{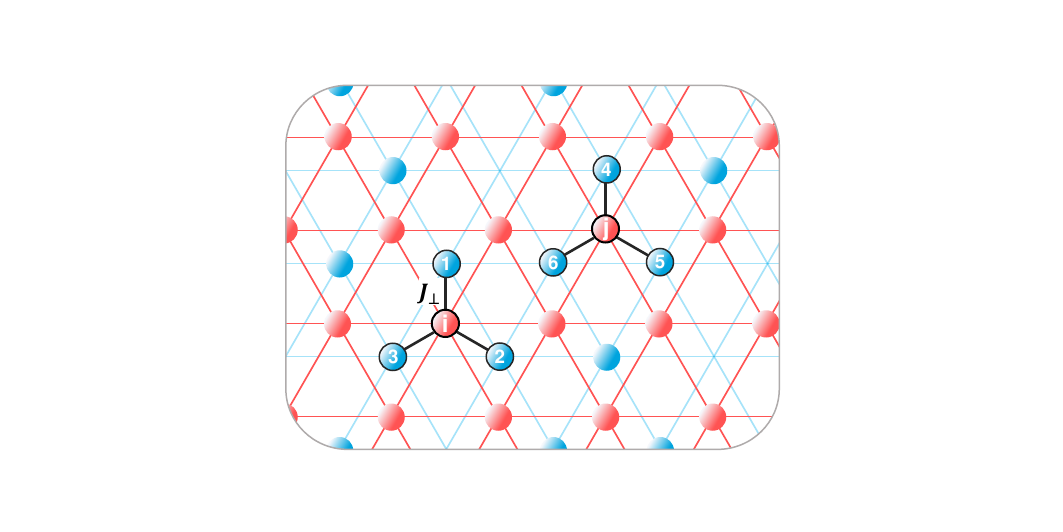}
\caption{\textbf{Bilayer Hubbard lattice and nomenclature of RKKY interactions.} Two local spins at sites $\mathbf{i}$ and $\mathbf{j}$ in the Mott insulating layer (red) interact by inducing a local magnetization profile in the metallic layer (blue). Each site in the Mott insulating layer has three nearest neighbors in the metallic layer, such that a total of nine terms contribute to Eq.~\eqref{eq:RKKY_int}.}
\label{fig:RKKY_fig}
\end{figure*} 

\end{document}


\title{Supplementary Information: \\ Implementation of the bilayer Hubbard model in a moir\'e heterostructure}

\author{Borislav Polovnikov*}
\author{Johannes Scherzer*}
\author{Subhradeep Misra*}
\author{Henning Schl\"omer}
\author{Julian Trapp}
\author{Xin Huang}
\author{Christian Mohl}
\author{Zhijie Li}
\author{Jonas G{\"o}ser}
\author{Jonathan F\"orste}
\author{Ismail Bilgin}
\author{Kenji Watanabe}
\author{Takashi Taniguchi}
\author{Annabelle Bohrdt}
\author{Fabian Grusdt}
\author{Anvar~S.~Baimuratov}
\author{Alexander H{\"o}gele}

\maketitle
\vspace{15pt}

\section{Supplementary Note 1: fabrication and experimental methods}
Monolayers of MoSe$_2$ and WS$_2$ were either mechanically exfoliated from bulk crystals (HQ Graphene) or obtained from in-house CVD synthesis. Thin flakes of hBN were exfoliated from bulk crystals provided by NIMS, Japan. All flakes were then stacked together by the dry-transfer technique using PDMS/PC stamps, deposited onto a Si/SiO$_2$ substrate and annealed at 200°C for 12 hours. The general sample design consisted of the MoSe$_2$/WS$_2$ heterobilayer (HBL) encapsulated in hexagonal boron nitride (hBN) and sandwiched between top and bottom few-layer graphene gates. When possible, we aimed at symmetric top and bottom gates in order to facilitate electrostatic control of the out-of-plane electric field and the charge density.

Five MoSe$_2$/WS$_2$ samples were studied in total, with two CVD-based H-type (antiparallel) samples, one CVD-based R-type (parallel) sample, and two exfoliated samples assembled from exfoliated flakes and identified as H-type by their spectroscopic signal. Since the CVD-grown flakes have a characteristic triangular shape, they were aligned a rotation angle of $180^\circ$ (H-type) or $60^\circ$ (R-type). In particular, sample S1 assembled from monolayers grown by chemical vapor deposition, and sample S2 was fabricated from exfoliated monolayers.

%
\begin{figure}[t!]  
\includegraphics[width=\textwidth]{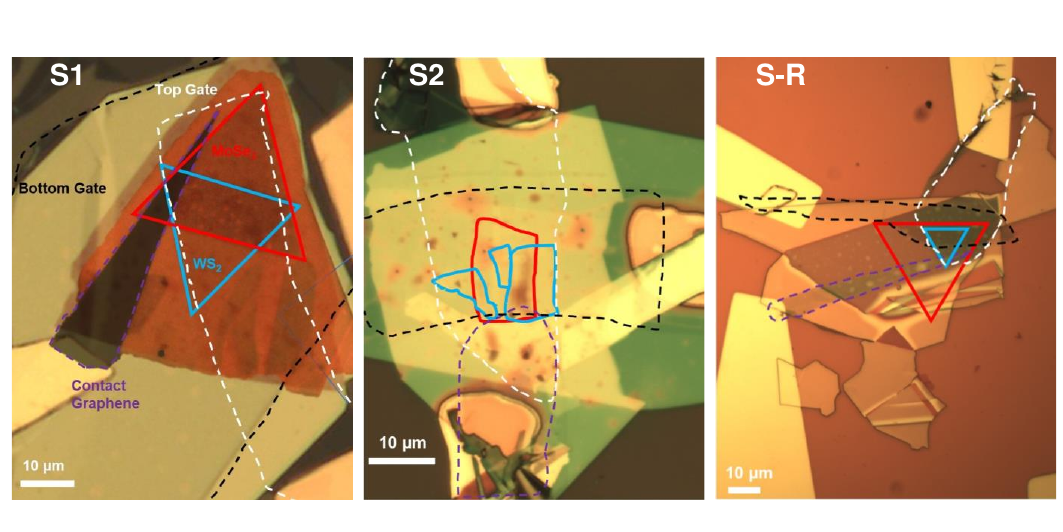}
\caption{Optical images of three studied MoSe$_2$/WS$_2$ devices. The left panel shows the CVD-based sample S1 that was already studied in Ref.~\cite{Polovnikov2023Apr}, the central panel shows the sample S2 assembled from exfoliated flakes, and the right panel shows the CVD-based R-type sample S-R.}
\label{assembly}
\end{figure}
%

The findings of this study are based on reflectance data obtained with home-built confocal microscopes in back-scattering geometry. The samples were loaded into a closed-cycle cryostat (attocube systems attoDRY1000) with a base temperature of $3.2$~K or into a dilution refrigerator (Leiden Cryogenics) with a base-temperature of $100$~mK and equipped with a superconducting magnet in Faraday configuration. Piezo-stepping and scanning units (attocube systems, ANPxyz and ANSxy100) were used for sample positioning with respect to a low-temperature apochromatic objective (attocube systems). 

Reflectance measurements were performed using a stabilized Tungsten-Halogen lamp (Thorlabs, SLS201L) or supercontinuum lasers (NKT Photonics, SuperK Extreme and SuperK Varia) as broadband light sources. The reflection signal was spectrally dispersed by monochromators (Roper Scientific, Acton SP2500 or Acton SpectraPro 300i with a 300 grooves/mm grating) and detected by liquid nitrogen or Peltier cooled charge-coupled devices (Roper Scientific, Spec-10:100BR or Andor, iDus 416). For control of the light polarization in excitation and detection, a set of linear polarizers (Thorlabs, LPVIS), half- and quarter-waveplates (B. Halle, $310-1100$~nm achromatic) mounted on piezo-rotators (attocube systems, ANR240) was used. Differential reflectance spectra were obtained by normalizing the reflected spectra from the HBL region ($R$) to that from the sample region without MoSe$_2$ and WS$_2$ layers ($R_0$) as $\textrm{DR} = (R-R_0)/R_0$. Additionally, polynomial background correction as detailed in Ref.~\cite{FengWang2019} was used when visualization of the data required a uniform signal dispersion.

\begin{figure}[h]  
\begin{center}
\includegraphics[width=1.0\textwidth]{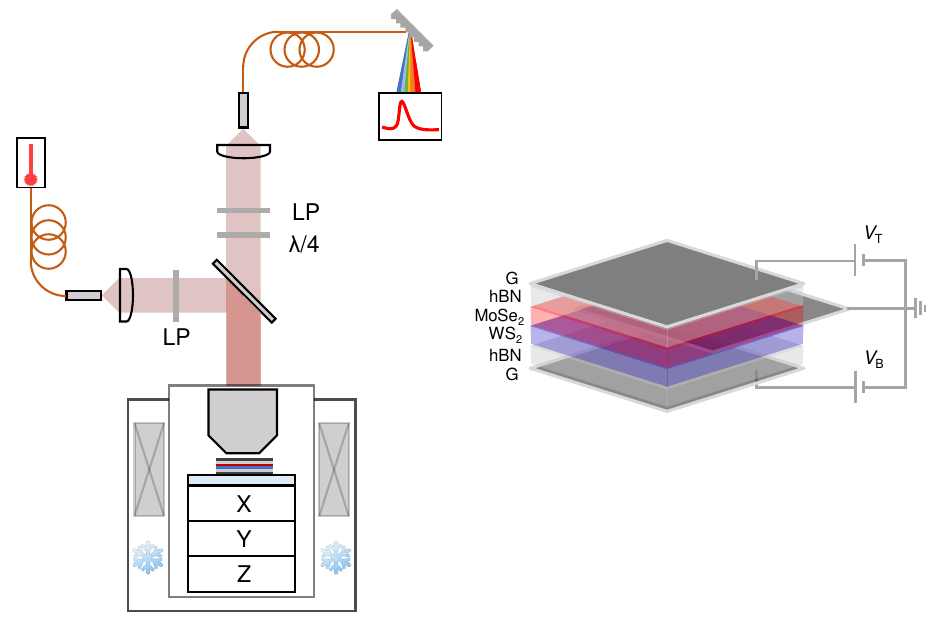}
\caption{Experimental setup. The left panel shows the confocal microscope with an excitation (horizontal) and a detection (vertical) arm mounted on top of a cryostat (attoDry1000 or Leiden Cryogenics Dilution Refrigerator). Inside the cryostat, Piezo positioners (attocube) were used to position the sample relative to a fixed low-temperature objective. The right panel shows the dual gate field-effect design employed throughout this study. Top- and bottom few-layer graphene gates were used to contorl electric field and charge doping, whereas insulating hexagonal boron nitride was used as the dielectric. This Figure was adapted from Ref.~\cite{Polovnikov2023Apr}.}
\label{Setup}
\end{center}
\end{figure}

\clearpage
\section{Supplementary Note 2: initial spectroscopic characterization}
\begin{figure*}[t!]  
\includegraphics[width=\textwidth]{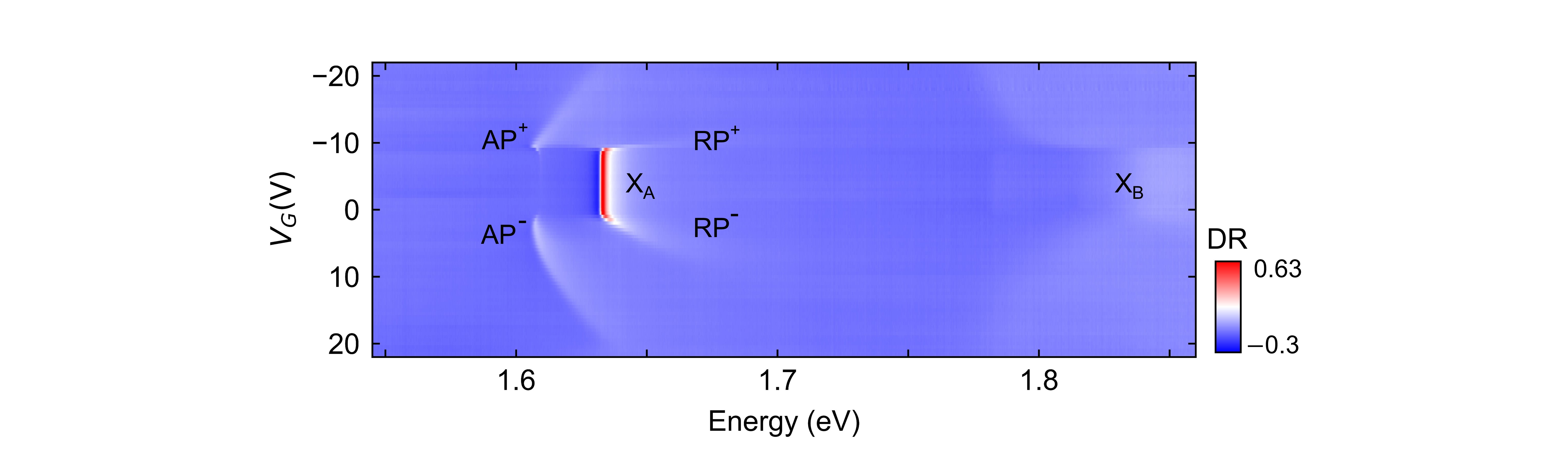}
\caption{Charge doping dependent DR signal showing ambipolar doping in monolayer MoSe$_2$ in sample S1. Changing the gate voltage $V_G=V_T=V_B$ from -22~V to 22~V results in a transition from a positive through the intrinsic to the electron doped regime with characteristic positive and negative attractive ($AP$) and repulsive polaron ($RP$) branches. The energy splitting of $28$~meV between the negative polaron branches ($RP^-$ and $AP^-$) of the A-exciton at the onset of electron doping is approximately 28~meV. Additionally, at around 1.85~eV, a faint peak that is interpreted as the B-exciton can be observed.}
\label{fig_charging_ML}
\end{figure*}

In Fig.~\ref{fig_charging_ML}, we show the evolution of DR in monolayer MoSe$_2$ in sample S1 as a function of the gate voltage $V_G=V_T=V_B$. Notably, the spectra show a transition from the p-doped (holes) through the intrinsic to the n-doped (electrons) regime, with characteristic signatures of the intralayer A-exciton $X_A$ and its attractive (AP) and repulsive polarons (RP). Additionally, the good quality of the sample is reflected by the relatively strong absorption by the intralayer B-exciton $X_B$ at around 1.85~eV. Consistent with the state of the literature, the charged attractive polarons of $X_B$ present a pronounced red-shift with increasing charge carrier density, whereas the attractive polarons of the fundamental $X_A$ exciton exhibit blue-shifts~\cite{Mak_Shan_WSe2_iX_2018}.

In all of our studied angle-aligned heterobilayer samples, both the absorption and the photoluminescence of WS$_2$ were strongly suppressed as illustrated for the sample S2 in Figure~\ref{ws2_signal} below. We are aware that some other studies~\cite{Ciorciaro2023May, Tang2022} have observed a weak WS$_2$ signal from the heterobilayer region, and as for now, we lack for decisive arguments to explain the difference between the observed behaviors. Ultimately, this contrast might be caused by the different spin-order of the sub-bands in the different stackings, with the H-type MoSe$_2$/WS$_2$ losing the whole WS$_2$ oscillator strength due to the energetically close 2s A- and 1s B-excitons.
%
\begin{figure}[t!]  
\begin{center}
\includegraphics[width=1.0\textwidth]{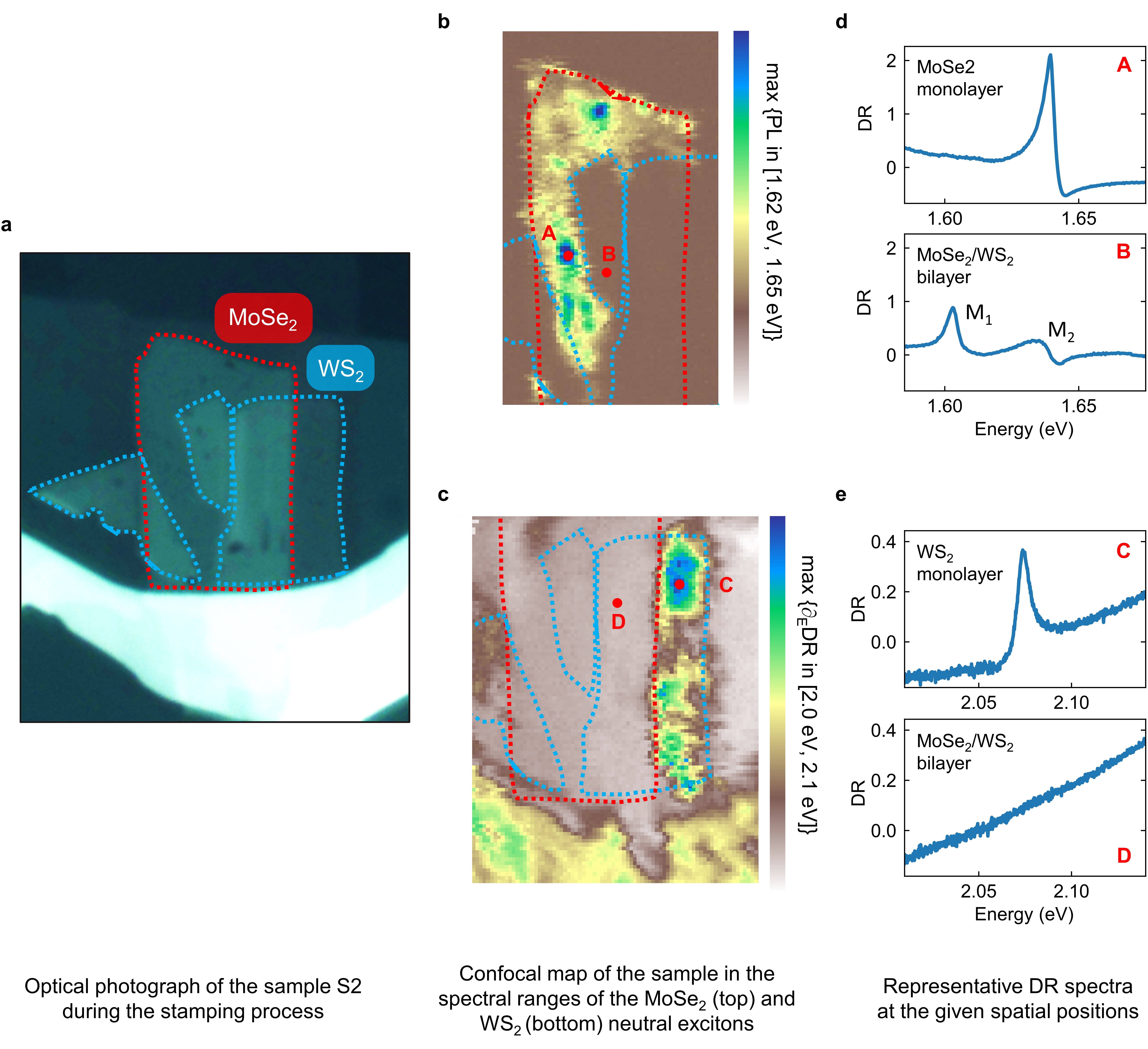}
\caption{(a) Optical photograph of the sample S2  where the red and blue dashed lines delimit the MoSe$_2$ and WS$_2$ monolayers, respectively. (b) Hyperspectral map of the sample acquired by projecting the maximum of the PL signal in the spectral range $1.62 - 1.65$~eV, with the bright areas reproducing the shape of the MoSe$_2$ monolayer. (c) Hyperspectral map of the sample acquired by projecting the maximum of the derivative of the DR signal in the spectral range $2.0 - 2.1$~eV, with the bright areas reproducing the shape of the WS$_2$ monolayer. (d) Representative DR spectra of the MoSe$_2$ monolayer and MoSe$_2$/WS$_2$ heterobilayer in the spectral range of the MoSe$_2$ intralayer exciton at the positions A and B shown in (b). (e) Representative DR spectra of the WS$_2$ monolayer and MoSe$_2$/WS$_2$ heterobilayer in the spectral range of the WS$_2$ intralayer exciton. Notably, no WS$_2$ transition is visible in the heterobilayer region.}
\label{ws2_signal}
\end{center}
\end{figure}

\begin{figure}[t!]  
\begin{center}
\includegraphics[width=0.9\textwidth]{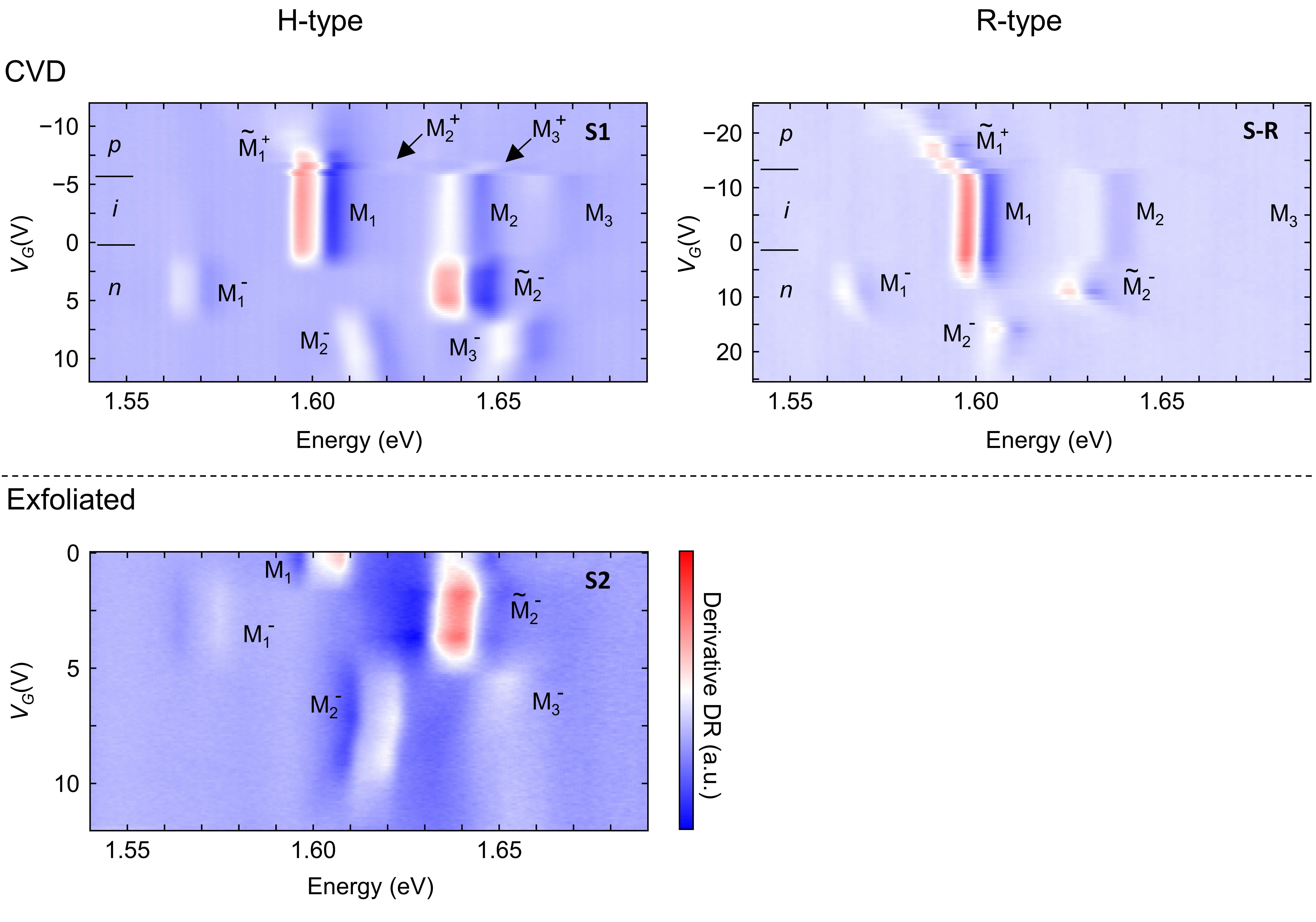}
\caption{Top: charge-doping dependent data in the CVD-based MoSe$_2$/WS$_2$ heterobilayers shown in Fig.~\ref{assembly}. In the H-type stacking (left) the peak $\tilde{M}_2^-$ is stabilized in an extended range of voltage, whereas in R-type (right) it presents a single maximum and pronounced blueshift away from this maximum at lower and higher gate voltages. Bottom: comparing the charge-doping dependent data of sample S2 with the two CVD-grown samples identifies its stacking regime as H-type.}
\label{assignment_stacking}
\end{center}
\end{figure}

We used the spectroscopic data of the CVD-based samples S1 and S-R (see Fig.~\ref{assembly}) in order to differentiate between H- and R-type stacking in the exfoliated samples. From our data on the CVD grown devices (Fig.~\ref{assignment_stacking}) it is visible that in the H-type stacking the peak $\tilde{M}_2^-$ is stabilized in an extended range of voltage, whereas it presents a single maximum and pronounced blueshift away from this maximum at lower and higher gate voltages in the R-type stacking. Furthermore, complemented by the data in Ref.~\cite{Ciorciaro2023May}, our data suggests that the peak that we call $M_3^-$ is bright in H-type and dark in R-type stacking. Both features can therefore be used to assign the stacking in the exfoliated devices as illustrated for the samples S2 in Fig.~\ref{assignment_stacking}.

%

We gain further insight into the nature of the neutral and charged excitons by probing their response to out-of-plane magnetic fields and measuring their g-factors. In linear response theory, an electron or hole interacts with the magnetic field through its spin and angular momentum. An electron (hole) with momentum $k$ changes its potential energy by a Zeeman shift $V_Z(k)$ proportional to $B$~\cite{Foerste2020,Wozniak2020},
\begin{equation}
V_Z(k) = \mu_B B \left[ g_0 s + L(k)\right] \, ,
\end{equation}
where $\mu_B = 57.9$~$\mu$eV/T is the Bohr magneton, $g_0 \simeq 2$ the free electron Land\'e factor, $s = \pm 1/2$ the (out-of-plane) spin of a given electronic state and $L(k)$ its angular momentum.  The opposite sign of $s$ and $L$ in the two $\pm K$ valleys (i.e. $L(K)=-L(-K)$ from time-revesal symmetry) implies that their Zeeman shift is also opposite. As a result, the excitonic resonance probed in either $\sigma^+$ or $\sigma^-$ polarized light will present an energy shift itself, with the valley Zeeman energy splitting given by
\begin{equation}\label{eq_2_3_g_factor}
\Delta E_Z = E(\sigma^+)-E(\sigma^-) \equiv g \mu_B B .
\end{equation}
In the last equation, we inroduced the exciton Land\'e factor $g=\Delta E_Z / (\mu_B B)$ as a dimensionless number characterizing the behavior of an exciton under out-of-plane magnetic fields. It measures the exciton dispersion as a function of $B$, and as such it can act both as a reporter on the electronic subbands involved in the exciton as well as on the local magnetization in the exciton's neighborhood.

\begin{figure*}[t!]  
\includegraphics[scale=1.0]{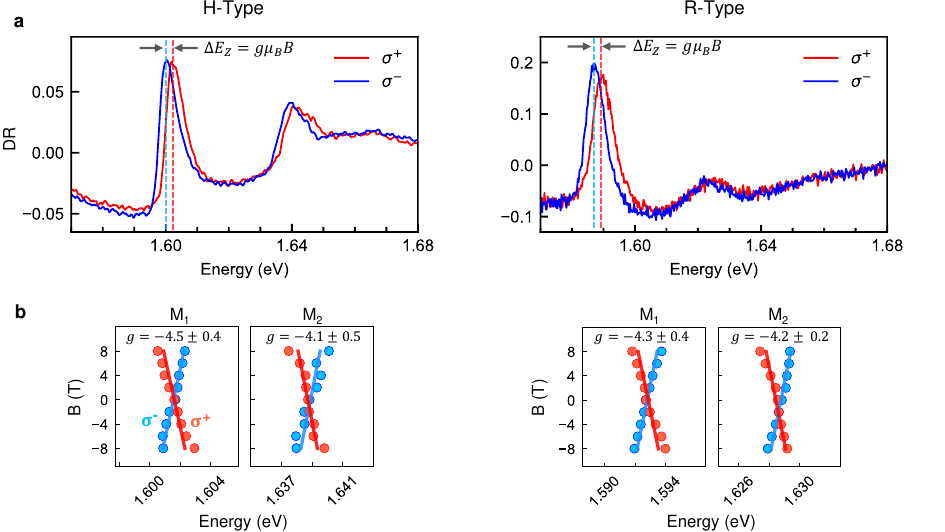}
\caption{(a) Charge-neutral DR spectra of the samples S1 (H-type) and S-R (R-type) in $\sigma^+$  and $\sigma^-$  polarization at an out-of-plane magnetic field of $B=-8$~T. (b) Evolution of the moir\'e exciton energies in $\sigma^\pm$ polarization as a function of the magnetic field. For both brightest peaks $M_1$ and $M_2$, the g-factors show values around the g-factor of the MoSe$_2$ intralayer A-exciton of around -4.}
\label{fig_magnetism1}
\end{figure*}

In Fig.~\ref{fig_magnetism1}a, we show the neutral DR spectra of the samples S1 (H-type) and S-R (R-type) in $\sigma^+$  and $\sigma^-$  polarization at an out-of-plane magnetic field of $-8$~T. The Zeeman splitting $\Delta E_Z$ that is visible for both the $M_1$ and $M_2$ resonances can be measured as function of the applied field B, resulting in the energetic dispersion illustrated in Fig.~\ref{fig_magnetism1}b. Notably, both of the bright moir\'e excitons show a g-factor close to the value of the MoSe$_2$ intralayer A-exciton around $-4$, confirming that they are of intralayer character themselves.

\begin{figure*}[t!]  
\includegraphics[scale=1.0]{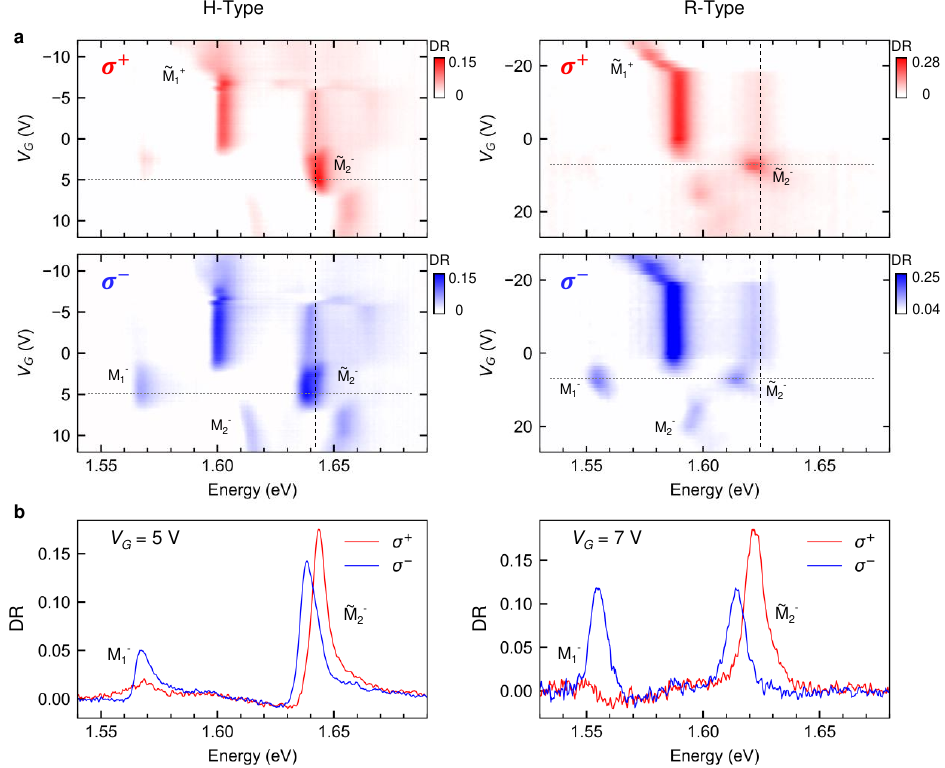}
\caption{(a) Doping-dependent DR spectra in $\sigma^+$ (top) and $\sigma^-$ (bottom) polarization at an out-of-plane magnetic field of $B=-8$~T. (b) Line cut of the DR-signal at $V_G = 5$~V (H-type) and $V_G = 7$~V (R-type) showing a polarized negatively charged trion $M_1^-$ and a pronounced Zeeman shift of $\tilde{M}_2^-$. }
\label{fig_magnetism2}
\end{figure*}

In Fig.~\ref{fig_magnetism2}, we show the doping-dependent absorption spectra in the two samples. At a perpendicular magnetic field of $-8$~T, the peak $M_1^-$ features almost complete valley polarization, just like Fermi polarons in monolayer MoSe$_2$ \cite{Back2017}. The peak $\tilde{M}_2^-$ (and similarly $\tilde{M}_1^+)$), on the other side, is only weakly polarized and shows a pronounced doping-dependent Zeeman splitting with nonlinear dependence on the magnetic field.

\begin{figure}[t!]  
\begin{center}
\includegraphics[scale=1.0]{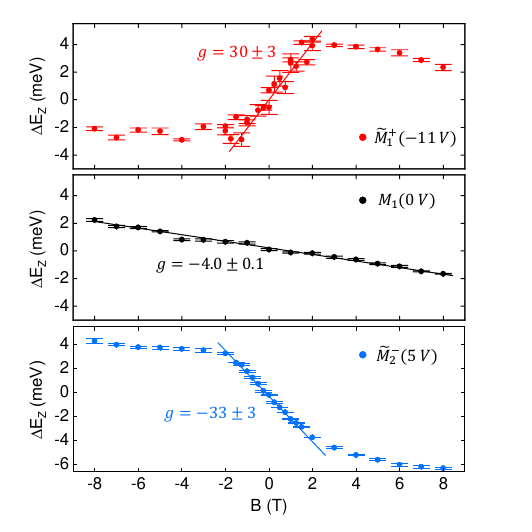}
\caption{Evolution of the valley Zeeman splitting $\Delta E_Z$ of the hole-doped peak $\tilde{M}_1^+$ (top) at $V_\text{G}=-11$~V, the neutral moir\'e peak $M_1$ (center) and the electron-doped peak $\tilde{M}_2^-$ (bottom) at $V_\text{G}=5$~V in device S1. The splittings of $\tilde{M}_1^+$ and $\tilde{M}_2^-$ depend non-linearly on the magnetic field, with a maximum $g$-factor of $30\pm3$  and $-33 \pm 3$, respectively (determined from linear fits for $B\lesssim 2$~T). The non-linear evolution of the $g$-factor is indicative of correlation-induced magnetism probed by the respective sensing exciton.}
\label{fig_magnetism3}
\end{center}
\end{figure}

To illustrate the dispersion of $\Delta E_Z$ as function of $B$, we show representative data on the p-doped, the intrinsic and the n-doped regimes in Fig.~\ref{fig_magnetism3}. The top panel shows the evolution of the Zeeman energy splitting of the p-doped peak $\tilde{M}_1^+$ which acts as a sensor exciton for the emerging hole lattice and shows a strongly renormalized, non-linear g-factor of $+30$. The central panel shows the corresponding evolution of $\Delta E_Z$ for the neutral moir\'e peak $M_1$ with the expected linear dispersion and a g-factor of $-4$. Finally, the bottom panel shows the Zeeman splittings of the n-doped peak $\tilde{M}_2^-$ which acts as a sensor exciton for the emerging electron lattice and also shows a strongly renormalized, non-linear g-factor of $-33$. We note that the magnetic response of $\tilde{M}_1^+$ and $\tilde{M}_2^-$ are in striking contrast: whereas in the latter case the g-factor is increased into the negative, for the former peak the g-factor changes its sign to positive values, consistent with other studies on hole-mediated spin lattices~\cite{Tang2020,Gerardot_exciton-polarons_2022}. This shows that the p-doped regime of MoSe$_2$/WS$_2$ also presents rich physics that has yet to be investigated in future studies, and the fact that one heterostructure provides simultaneous access to both electron and hole-doped correlated states makes it standing out in the field.
%

\newpage
\clearpage
\section{Supplementary note 3: modeling of moiré excitons}

\begin{figure}[h!]  
\begin{center}
\includegraphics[width=1.0\textwidth]{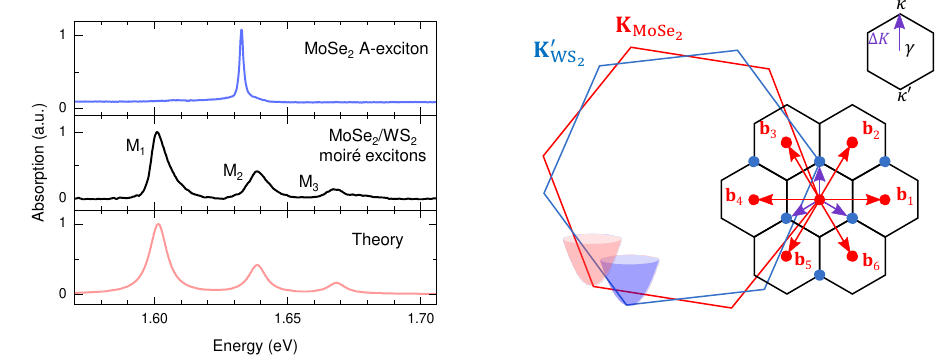}
\caption{Left panel, from top to bottom: optical absorption of the monolayer MoSe$_2$ A-exciton in sample S1, the optical absorption of the three moir\'e excitons $M_1$, $M_2$ and $M_3$ in the bilayer region of the same sample, and the theoretical fit of the three moir\'e excitons. Right panel: momentum space geometry of angle aligned H-type MoSe$_2$/WS$_2$. The momentum mismatch $\Delta K$ between the K-valley of MoSe$_2$ and the K'-valley of WS$_2$ defines the mini Brillouin Zone that reflects the long-range periodicity of the moir\'e superlattice. This Figure was adapted from Ref.~\cite{Polovnikov2023Apr}.}
\label{moire_theory}
\end{center}
\end{figure}

To model the moir\'e excitons in MoSe$_2$/WS$_2$ structures, we employ a continuum model that is based on the models introduced in Refs.~\cite{MacdonaldTopo2017,MacdonaldIX2018,MacdonaldHubbard2018,RuizFalko2019} and has been presented in detail in Ref.~\cite{Polovnikov2023Apr}. We assume that the two monolayers are angle-aligned, such that either both K-valleys (R-type) or the K-valley of MoSe$_2$ and the K'-valley of WS$_2$ (H-type) are closely aligned. The right panel of Figure~\ref{moire_theory} shows the geometry of the mini Brillouin Zone (mBZ) which we define such that the center of the mBZ $\gamma$ overlaps with the K-valley of MoSe$_2$.

The left panel of Figure~\ref{moire_theory} shows the optical absorption by the A-exciton of the monolayer MoSe$_2$ region in sample S1 as well as the three bright moir\'e excitons in the heterobilayer region of the same sample. Since the optical response of the device is entirely dominated by the MoSe$_2$ valence bands, we limit our discussion to the momentum bright exciton states in the center of the mBZ and assume parabolic exciton dispersion, i.e. $E(\mathbf{k}) = E_\text{X} + \hbar^2 |\mathbf{k}|^2 / (2M_\text{X})$ and $\mathcal{E}(\mathbf{k'}) = E_\text{IX} + \hbar^2 |\mathbf{k'}|^2 / (2M_\text{IX})$ for intra- and interlayer excitons, respectively.

The long-range order introduced by the moir\'e potential introduces scattering between states with a momentum mismatch given by the moir\'e reciprocal vectors $\mathbf{b}_i$ ($i=1,...,6$) as denoted in Figure~\ref{moire_theory}. Here, we consider the first six intralayer states as marked by the red dots, and following Refs.~\cite{MacdonaldTopo2017,MacdonaldIX2018,MacdonaldHubbard2018} we introduce a phenomenological moir\'e potential $V(\mathbf{r})=\sum_{j=1}^6 V_j \exp{(i\mathbf{b}_j\mathbf{r})}$ as a scattering potential. In particular, the 120$^\circ$ symmetry enforces $V_1 = V_3 = V_5 \equiv V$ and $V_2 = V_4 = V_6 \equiv V^*$ as elaborated in Refs.~\cite{MacdonaldTopo2017,MacdonaldIX2018,MacdonaldHubbard2018}. Further, we pay heed to the presence of WS$_2$ bands by introducing an interlayer hopping potential $t$ that couples the intralayer states (red dots) with the three closest interlayer states (blue dots) in momentum space as specified by the violet arrows in Fig.~\ref{moire_theory}~\cite{Yu_IXcoupling_2017,RuizFalko2019}.

%

The total Hamiltonian~\ref{HamMatrix} mixes seven intralayer and six interlayer states into a new set of eigenstates that we call moir\'e excitons. It is composed of blocks of intralayer exciton scattering (red diagonal block with $E_j(\mathbf{k}) = E (\mathbf{k-b}_i)$ for $i=1,...,6$) and interlayer exciton states (blue diagonal block with $\mathcal{E}_\eta(\mathbf{k}) = \mathcal{E}(\mathbf{k} - C^\eta_3 \Delta \mathbf{K})$ and $\mathcal{E}_\zeta(\mathbf{k}) = \mathcal{E}(\mathbf{k} + 2 C^{\zeta-3}_3 \Delta \mathbf{K})$ for $\eta=0,1,2$ and $\zeta=3,4,5$). Finally, hermitian coupling of the two sets of exciton states by the off-diagonal blocks with hopping elements $t$ complete the model and define the Hamiltonian as:

\begin{equation}\label{HamMatrix}
\arraycolsep = 0.2pt
\def\arraystretch{.5}
\footnotesize
    H(\mathbf{k}) = 
  \left(\begin{tblr}{
    colspec = {lllllllllllll},
    cell{1}{1,2,3,4,5,6,7} = {red!6},
    cell{2}{1,2,3,4,5,6,7} = {red!6},
    cell{3}{1,2,3,4,5,6,7} = {red!6},
    cell{4}{1,2,3,4,5,6,7} = {red!6},
    cell{5}{1,2,3,4,5,6,7} = {red!6},
    cell{6}{1,2,3,4,5,6,7} = {red!6},
    cell{7}{1,2,3,4,5,6,7} = {red!6},
    cell{8}{8,9,10,11,12,13} = {blue!6},
    cell{9}{8,9,10,11,12,13} = {blue!6},
    cell{10}{8,9,10,11,12,13} = {blue!6},
    cell{11}{8,9,10,11,12,13} = {blue!6},
    cell{12}{8,9,10,11,12,13} = {blue!6},
    cell{13}{8,9,10,11,12,13} = {blue!6},
  }
    E_0  & V     & V^* & V   & V^* & V   & V^* & t    & t    & t    & 0    & 0    & 0 \\
    V^* & E_1  & V   & 0   & 0   & 0   & V   & 0    & 0    & t    & 0    & t    & 0 \\
    V     & V^* & E_2 & V^* & 0   & 0   & 0   & t    & 0    & 0    & 0    & t    & 0 \\
    V^* & 0     & V   & E_3 & V   & 0   & 0   & t    & 0    & 0    & 0    & 0    & t \\
    V     & 0     & 0   & V^* & E_4 & V^* & 0   & 0    & t    & 0    & 0    & 0    & t \\
    V^* & 0     & 0   & 0   & V   & E_5 & V   & 0    & t    & 0    & t    & 0    & 0 \\
    V     & V^* & 0   & 0   & 0   & V^* & E_6 & 0    & 0    & t    & t    & 0    & 0 \\
    t^*  & 0  & t^* & t^* & 0   & 0   & 0   & \mathcal{E}_0 & 0  & 0 & 0 & 0 & 0 \\
    t^*  & 0   & 0 & 0 & t^* & t^* & 0 & 0  & \mathcal{E}_1 & 0    & 0    & 0    & 0 \\
    t^*  & t^* & 0   & 0   & 0  & 0   & t^* & 0  & 0  & \mathcal{E}_2 & 0    & 0    & 0 \\
    0    & 0    & 0  & 0  & 0 & t^* & t^* & 0  & 0  & 0  & \mathcal{E}_3 & 0    & 0 \\
    0   & t^* & t^* & 0  & 0   & 0  & 0  & 0   & 0   & 0   & 0   & \mathcal{E}_4 & 0 \\
    0   & 0   & 0 & t^* & t^* & 0  & 0  & 0   & 0   & 0   & 0   & 0   & \mathcal{E}_5 
  \end{tblr}\right) .
\end{equation}
\\
Finally we emphasize that this model can only be fitted by considering field-dependent exciton dispersion as discussed in Ref.~\cite{Polovnikov2023Apr}. Otherwise, the influence of the intralayer potential $V$ and the interlayer tunneling $t$ on the moir\'e exciton mixing can not be disentangled.

\clearpage